\documentclass[sigconf]{acmart}

\renewcommand\footnotetextcopyrightpermission[1]{} % removes footnote with conference information in first column
\usepackage{booktabs} % For formal tables

\usepackage{float}
\usepackage{graphicx}
\usepackage{amsmath}
\usepackage[titlenumbered,linesnumbered,ruled,algonl,vlined,noresetcount]{algorithm2e}
\usepackage{array}
\usepackage{balance}
\usepackage[format=hang,singlelinecheck=on]{caption}

\newcommand{\lrg}[1]{\large\emph#1\small}

\newcommand{\func}[2]{
	\SetKwFunction{FMain}{#1}
	\SetKwProg{Fn}{Function}{:}{}
	\Fn{\FMain{#2}}
}

\begin{document}
\title{The Optimal Route and Stops for a Group of Users\\ in a Road Network}
%\titlenote{Produces the permission block, and
  %copyright information}
%\subtitle{Extended Abstract}
%\subtitlenote{The full version of the author's guide is available as
%  \texttt{acmart.pdf} document}

\author{Radi Muhammad Reza}
%\authornote{Dr.~Trovato insisted his name be first.}
\affiliation{%
  \institution{Department of CSE, BUET}
  \streetaddress{Dhaka-1000, Bangladesh}
  %\city{Dhaka} 
}
\email{radireza@gmail.com}

\author{Mohammed Eunus Ali}
%\authornote{The secretary disavows any knowledge of this author's actions.}
\affiliation{%
  \institution{Department of CSE, BUET}
  \streetaddress{Dhaka-1000, Bangladesh}
}
\email{eunus@cse.buet.ac.bd}

\author{Muhammad Aamir Cheema}
%\authornote{The secretary disavows any knowledge of this author's actions.}
\affiliation{%
  \institution{Faculty of IT, Monash University}
  \streetaddress{Melbourne, Australia}
}
\email{aamir.cheema@monash.edu}

\begin{abstract}
%Recently, with the advancement of the GPS-enabled cellular technologies, the location-based services (LBS) have gained in popularity.  Nowadays, an increasingly larger number of map-based applications enable users to ask a wider variety of queries. Researchers have studied the ride-sharing, the carpooling, the vehicle routing, and the collective travel planning problems extensively in recent years. Collective traveling has the benefit of being environment-friendly by reducing the global travel cost, the greenhouse gas emission, and the energy consumption. 

Recently, with the advancement of the GPS-enabled cellular technologies, the location-based services (LBS) have gained in popularity.  Nowadays, an increasingly larger number of map-based applications enable users to ask a wider variety of queries. Researchers have studied the ride-sharing, the carpooling, the vehicle routing, and the collective travel planning problems extensively in recent years. Collective traveling has the benefit of being environment-friendly by reducing the global travel cost, the greenhouse gas emission, and the energy consumption. In this paper, we introduce several optimization problems to recommend a suitable route and stops of a vehicle, in a road network, for a group of users intending to travel collectively. The goal of each problem is to minimize the aggregate cost of the individual travelers' paths and the shared route under various constraints. First, we formulate the problem of determining the optimal pair of end-stops, given a set of queries that originate and terminate near the two prospective end regions. We outline a baseline polynomial-time algorithm and propose a new faster solution - both calculating an exact answer. In our approach, we utilize the path-coherence property of road networks to develop an efficient algorithm. Second, we define the problem of calculating the optimal route and intermediate stops of a vehicle that picks up and drops off passengers en-route, given its start and end stoppages, and a set of path queries from users. We outline an exact solution of both time and space complexities exponential in the number of queries. Then, we propose a novel polynomial-time-and-space heuristic algorithm that performs reasonably well in practice. We also analyze several variants of this problem under different constraints. Last, we perform extensive experiments that demonstrate the efficiency and accuracy of our algorithms.
\end{abstract}

%\begin{CCSXML}
%<ccs2012>
 %<concept>
  %<concept_id>10010520.10010553.10010562</concept_id>
  %<concept_desc>Computer systems organization~Embedded systems</concept_desc>
  %<concept_significance>500</concept_significance>
 %</concept>
 %<concept>
  %<concept_id>10010520.10010575.10010755</concept_id>
  %<concept_desc>Computer systems organization~Redundancy</concept_desc>
  %<concept_significance>300</concept_significance>
 %</concept>
 %<concept>
  %<concept_id>10010520.10010553.10010554</concept_id>
  %<concept_desc>Computer systems organization~Robotics</concept_desc>
  %<concept_significance>100</concept_significance>
 %</concept>
 %<concept>
  %<concept_id>10003033.10003083.10003095</concept_id>
  %<concept_desc>Networks~Network reliability</concept_desc>
  %<concept_significance>100</concept_significance>
 %</concept>
%</ccs2012>  
%\end{CCSXML}

%\ccsdesc[500]{Computer systems organization~Embedded systems}

%\keywords{ACM proceedings, \LaTeX, text tagging}

\maketitle

\section{Introduction}
\label{intro}

The proliferation of the GPS-equipped cellular devices and the map-based applications have enabled people to obtain their location data and other spatial information instantly. The location-based services (LBS) use this information to solve a variety of queries. Nowadays, everyone expects to find a suitable LBS to answer any travel related query s/he may feel the need to ask. In this paper, we formulate and investigate a range of new queries that facilitate collective traveling of a group of users using a single vehicle.

\begin{figure}[!tb]
\centering
\includegraphics[width=\columnwidth]{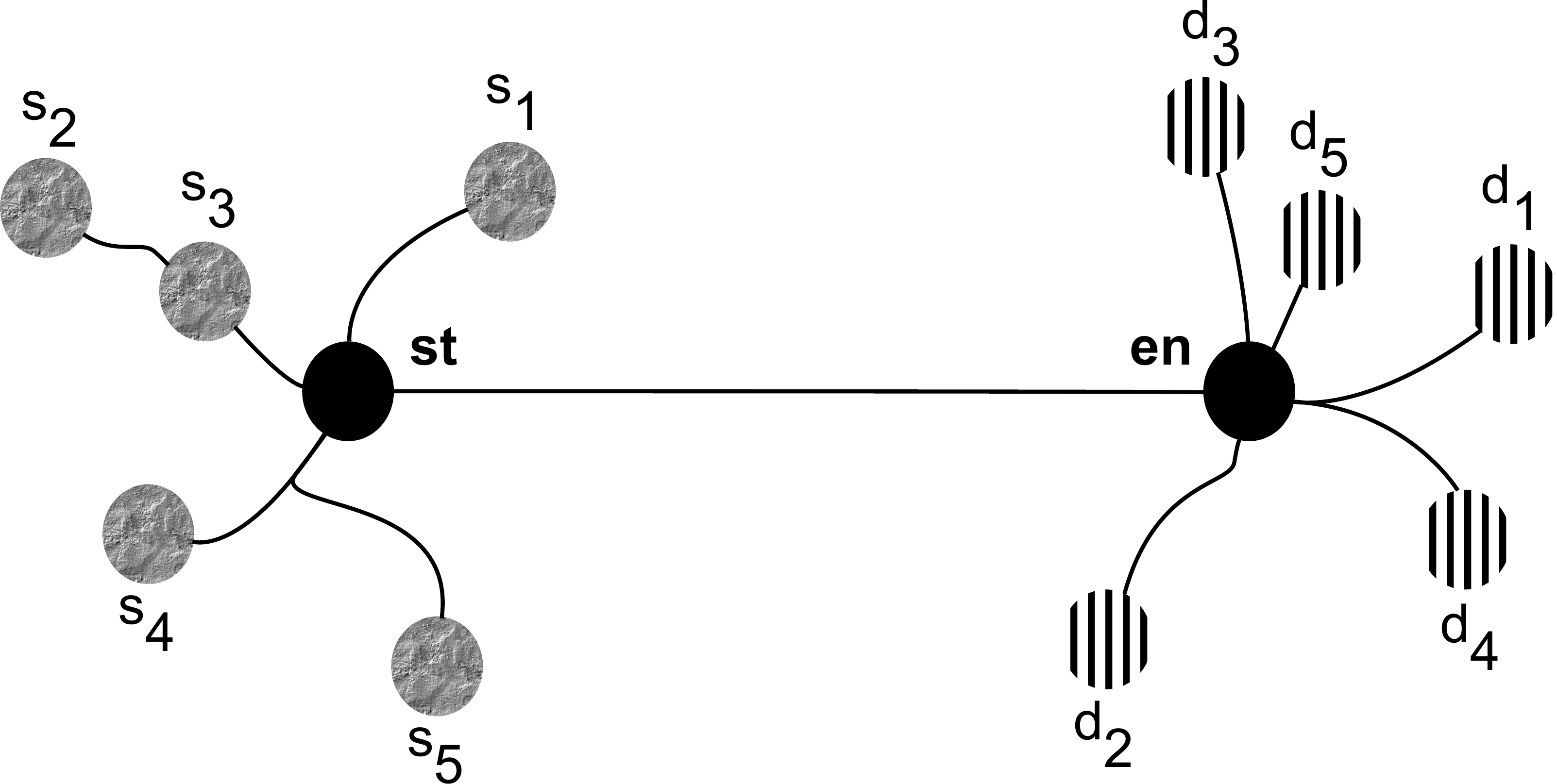}
\caption{The OES Query.}
\label{fig_oes}
\end{figure}

In our first problem, we determine the optimal start-and-end-stops of a vehicle, given path queries from co-located sources to co-located destinations; the vehicle picks all passengers up from its start-stop and drops them off at its end-stop. We name this problem as the optimal end-stops $(OES)$ query. In Figure~\ref{fig_oes}, the source nodes are co-located in a region, while the destination nodes are co-located in a distant region. The goal is to determine an optimal pair of end-stops $(st,en)$, which minimizes the summation of the shortest path cost between $st$ and $en$, the travel costs from $s_{i}$s to $st$, and the costs from $en$ to $d_{i}$s.

A demand-based transportation agency that provides vehicles to carry people across a city/state and assigns a group of passengers to a particular vehicle may use the $OES$ query to determine the vehicle's optimal end points. Vehicular service for tourists traveling from one hot-spot to another or friends planning a picnic may also benefit from this query by help determining the optimal meeting location. Any group of people desiring to travel collectively may decide upon the gathering, and the disperse points by using this query.

\begin{figure}[!tb]
\centering
\includegraphics[width=\columnwidth]{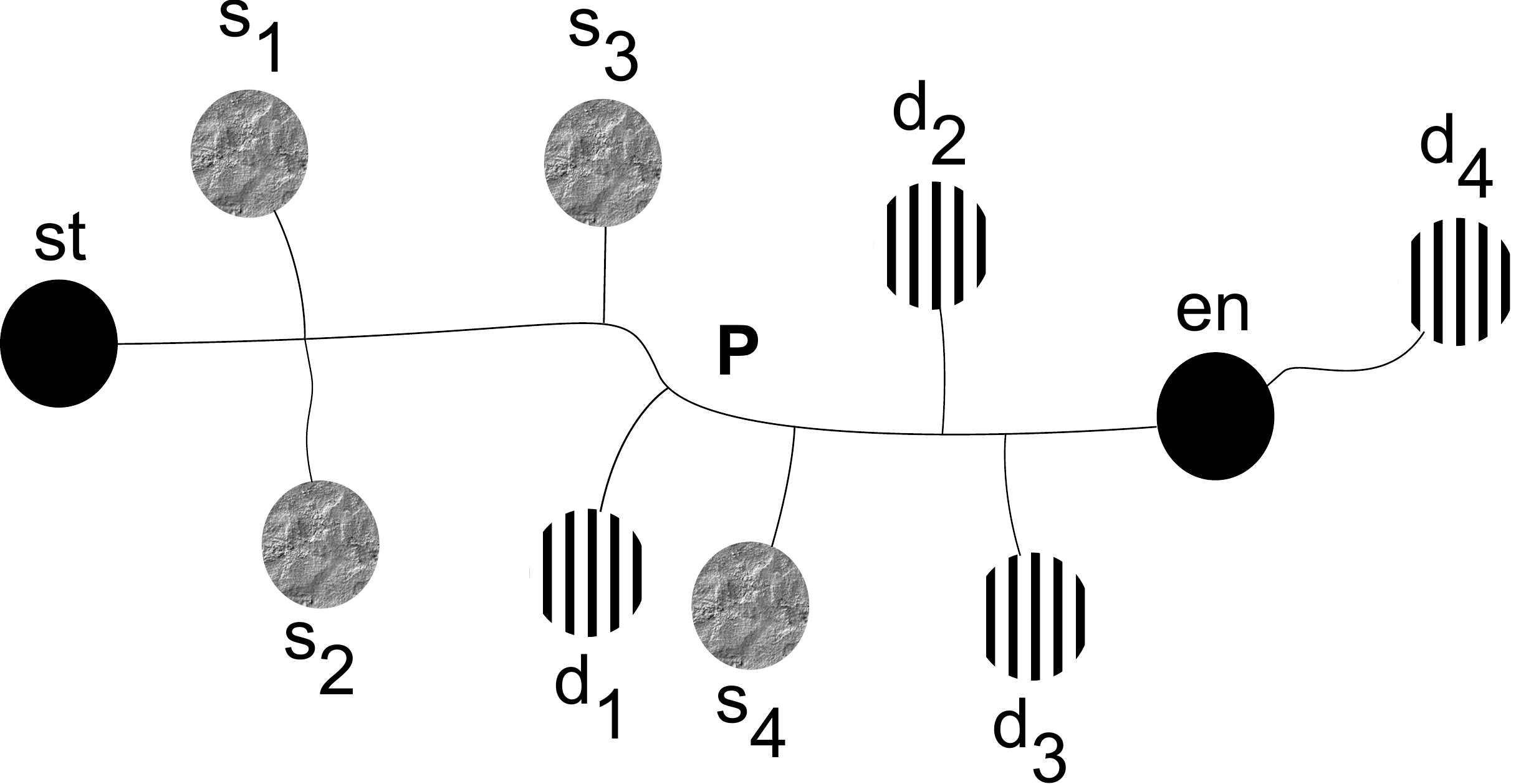}
\caption{The ORIS Query.}
\label{fig_oris}
\end{figure}

Our second problem is to determine the optimal route and the intermediate pick-up and drop-off locations along the path of a vehicle, given its two end-stops and query sources and destinations near its potential route. We call this problem as the optimal route and intermediate stops $(ORIS)$ query. In Figure~\ref{fig_oris}, the query nodes are in locations that make sense. The objective is to compute an optimal route $P$ from $st$ to $en$, which minimizes the summation of the cost of $P$, the costs from $s_{i}$s to $P$, and the costs from $P$ to $d_{i}$s.

An off-campus bus service for an educational institution may ask our $ORIS$ query to determine its route and the locations to pick up and drop off students. A transportation service for office staffs may similarly benefit from this query. A person may want to pick friends up along the way to a restaurant, or a theater; s/he would also require determining her/his travel route and pick-up locations. In general, any vehicle with advance passenger-reservations, or any group of people planning to travel collectively by sharing a vehicle may ask this query to plan the optimally shared route in advance. Similarly, a ride-sharing system \cite{IDM-LAB:journals/elsevier/Koen13i}, after matching its passengers to a fleet of vehicles, may use this query to determine the optimal route and stops of each vehicle in the system. Again, a cargo transportation system with one heavy carrier shipping across cities, and several light carriers loading from/to the main vehicle may also use this kind of query to determine the transaction locations. The motivation behind our second query is to reduce the global travel cost of the main vehicle and the individual travelers' transportation to and from it; this reduction results in the diminution of the net energy consumption and an overall greener transport.

We introduce several variants of the $ORIS$ problem under different constraints. First, a user may not want to travel too far to get on/off the vehicle. S/He usually prefers the entry (resp. exit) point within walking distance of her/his source (resp. destination). Therefore in a variant, we constrain the maximum allowable path length of a user to or from the vehicle. Second, the vehicle's agency may want to limit its path length. For a reasonably large number of passengers, the optimal route - as calculated by our algorithm - tends to loop again and again to pick everyone from home and drop him/her at the destination. Such a looping path is unrealistic under practical considerations. Again, the driver does not want to run out of gas. Thus, limiting the maximum allowable path-length of the vehicle to formulate another variant is pragmatic. Third, for a large number of users, a driver may not want to stop to pick or drop a single person. If a large vehicle stops too often, it may inconvenience onboard passengers. To address this issue, we propose two variants - one by restricting the minimum number of people required to get on/off at a stopping point, another by directly limiting the allowable number of stops of the vehicle. Last, the vehicle's agency may need to assign unequal weights to the cost of the vehicle's route, and the total cost of the solo travel by the passengers. For a small number of queries, it is pragmatic to pick each user up from his/her source and drop him/her off at his/her destination; placing a small weight on the vehicle's path cost ensures the computation of such a route. On the other hand, a larger weight on the cost of the vehicle prevents a looping of its path and forces it to travel in a shorter route; this becomes practical, when the number of users increases. Besides these variants, we consider the possibility of deriving additional ones by adding more than one constraint at a time.

The path queries input to our problems and variants may come from two different types of LBSs. One treats each of our queries as a standalone application. It depends on either advanced booking of passengers, or joint planning by users to produce the query source-destination pairs. Then, it executes our algorithm offline on the generated set of path queries. The other type pipes the output of an existing clustering algorithm such as \cite{DBLP:conf/ssd/MahmudAAHN13}, or a vehicle-passenger matching algorithm of a ride-sharing system, e.g., \cite{DBLP:journals/pvldb/HuangBJW14}, \cite{DBLP:journals/tkde/MaZW15}, \cite{DBLP:conf/icde/MaZW13}, and \cite{DBLP:conf/gis/AlarabiCZMB16}, as input to each of our problems. Then, each run of our algorithm computes the best route of a vehicle for the passengers assigned to it. Ride-sharing has the benefit of saving time, money, and the environment \cite{DBLP:conf/percom/Stach11}, \cite{DBLP:conf/hicss/LiHZ17}, \cite{DBLP:conf/huc/CiciMFL14}, \cite{DBLP:conf/amcis/WangAS17}, \cite{Alonso-Mora17012017}. Our techniques have the potential to perform as the last stage in the pipeline of a ride-sharing system \cite{IDM-LAB:journals/elsevier/Koen13i}.

Besides ride-sharing \cite{IDM-LAB:journals/elsevier/Koen13i}, our novel queries relate to several other fascinating problems in the literature. The $OES$ problem is similar to, yet different from, the optimal meeting point $(OMP)$ query \cite{DBLP:journals/pvldb/YanZN11}, \cite{DBLP:journals/kais/YanZN15}. Given a set of query points, the $OMP$ query finds a gathering location that minimizes the aggregate travel cost of the users. This query does not take into account the location and orientation of two distant clusters of query nodes. Thus, the techniques to solve the $OMP$ problem cannot answer our $OES$ query. The other associated problems are the group nearest neighbor $(GNN)$ \cite{DBLP:conf/icde/PapadiasSTM04}, the group k-nearest neighbors $(GKNN)$ \cite{DBLP:journals/jgs/Safar08}, the k-optimal meeting points $(k-OMP)$ \cite{DBLP:conf/dnis/TiwariK13a}, and the optimal location $(OL)$ \cite{DBLP:reference/gis/Xiao17} queries in the road network. However, the solution methodologies for these problems are not applicable in solving the $OES$ query.

The $ORIS$ problem is a generalization of the traveling salesman path problem $(TSPP)$ \cite{DBLP:journals/mp/LamN08}. Our $ORIS$ query also relates to the trip planning $(TP)$ \cite{DBLP:conf/ssd/LiCHKT05}, \cite{DBLP:conf/ssd/HashemHAK13}, the optimal sequenced route $(OSR)$ \cite{DBLP:journals/vldb/SharifzadehKS08}, \cite{DBLP:conf/gis/CostaNMTPM15}, \cite{DBLP:conf/mdm/SamroseHBAUM15}, \cite{DBLP:journals/geoinformatica/ChenKSZ11}, \cite{DBLP:journals/geoinformatica/SharifzadehS08}, the keyword-aware optimal route $(KOR)$ \cite{DBLP:conf/icde/CaoCCGPX13}, the carpooling \cite{DBLP:conf/kdd/GeLXC11}, \cite{DBLP:conf/sensys/ZhangLZLLH13}, \cite{DBLP:journals/tetc/ZhangHLLS14}, the vehicle routing $(VRP)$ \cite{DBLP:journals/corr/Munari16}, \cite{DBLP:journals/candie/BraekersRN16}, \cite{DBLP:journals/corr/MahmoudiZ15}, \cite{DBLP:journals/dam/TothV02}, the collective travel planning $(CTP)$ \cite{DBLP:journals/tkde/ShangCWJWK16}, and the Steiner diagram \cite{DBLP:journals/jda/BlasumHOW07}, \cite{DBLP:reference/algo/Chuzhoy08}, \cite{DBLP:journals/networks/DreyfusW71} problems. Detour ride-sharing is another related, yet different, problem \cite{DBLP:conf/socs/DrewsL13}, \cite{DBLP:conf/atmos/GeisbergerLNSV10}. Be that as it may, none of these approaches are adequate in solving the $ORIS$ query. The optimal multi-meeting-point route search $(OMMPR)$ query \cite{DBLP:journals/tkde/LiQYM16} is very similar to our $ORIS$ problem. However, it only solves a variant of our problem that we introduce in Section~\ref{pc5} - it aims at minimizing a weighted sum of the vehicle's route cost and the users' query-node-to-route costs. Furthermore, it provides four dynamic programming algorithms, each of which has time and space complexities exponential in the number of queries. We develop a novel polynomial-time-and-space heuristic solution that computes an answer instantly, incurring a very low error. Unlike the algorithms in \cite{DBLP:journals/tkde/LiQYM16}, which work for a maximum of five users, our technique scales well for a large number of, like fifty, queries. We also introduce several variants, including the weighted version \cite{DBLP:journals/tkde/LiQYM16}, and propose modifications of our approach to solve those versions.

The shortest path computation problems, e.g., \cite{DBLP:journals/tcs/JuknaS16}, \cite{DBLP:books/daglib/0023376}, \cite{DBLP:books/lib/RussellN03}, \cite{DBLP:conf/ssd/ShekharFG97}, \cite{DBLP:conf/wea/GeisbergerSSD08}, and \cite{DBLP:conf/sigmod/ZhuMXLTZ13}, do not apply in our context. Yet, we benefit from the intuition behind the $Dijkstra$'s algorithm \cite{DBLP:books/daglib/0023376}, \cite{DBLP:conf/anziis/EklundKP96}, the bidirectional search \cite{DBLP:conf/aaai/DavisPS84}, and the group shortest path approach \cite{DBLP:conf/mdm/RezaAH15} in solving both the $OES$ and the $ORIS$ queries. Below, we provide the intuition, and a high-level overview of our solution strategies for the $OES$ and the $ORIS$ problems.

To solve the $OES$ problem, at first, we outline a baseline brute-force technique. The straightforward solution approach does not exploit the properties of road networks; thus, it is slow and inefficient. Then, we provide a more efficient algorithm. We adopt a simultaneous search technique that utilizes the path coherence property of road networks to improve the expected time complexity of our algorithm. In practical scenarios, our approach is many times faster than the baseline solution.

For the $ORIS$ problem, we first provide an exact solution of both time and space complexities exponential in the number of sources-and-destinations. The algorithm to compute an optimal answer works for only a small number of users. For practical purpose, we require a solution approach, which is scalable for a large number of passengers. To that end, we propose a polynomial-time-and-space heuristic algorithm that works well in practice. In this approach, instead of exploring an exponential number of sub-problems, we make greedy choices to keep the size of our search space within polynomial bound. Thus, we achieve an algorithm of polynomial time and space complexities; however, this gain in scalability comes at the price of a little accuracy. Our heuristic efficiently calculates a near-optimal answer for a large group of people with a reasonably small error. To solve each variant of our $ORIS$ query, we propose modifications of the above algorithms.

We perform extensive experiments on a real road network dataset, \cite{DBLP:journals/geoinformatica/Brinkhoff02}, using synthetic query pairs. Our experimental results demonstrate the efficiency and effectiveness of our solutions. Our first algorithm is an order of magnitude faster than the straightforward solution and still, produces an exact answer. Our approaches to the second problem and its variants incur an average relative error of less than 5\%; however, contrary to the exponential-time theoretical approaches, they produce an answer within a second. Our algorithms also save space as they have polynomial space complexities.

In summary, this paper has the following key points:
\begin{itemize}
\item We propose a novel type of query to determine the optimal end-stoppages of a vehicle for a group of users sharing it between co-located sources and destinations.
\item We provide a fast and exact solution to compute the end-stops, which outperforms the straightforward solution in practice.
\item We introduce another query to determine the optimal route and the intermediate stops of a vehicle for a batch of passengers entering/exiting the vehicle en-route; we formulate several variants of this problem under different constraints.
\item We propose novel polynomial-time heuristics for the vehicle-route-and-stops problem and its variants; our heuristics are far more efficient than exponential-time exact solutions and produce a near-optimal answer.
\item We conduct extensive experiments to investigate the efficiency and effectiveness of the developed algorithms.
\end{itemize}

\section{Problems Formulation}
\label{prob}
In this section, we introduce our problems of determining the optimal route and the stops of a vehicle in a road network. In each problem, we are given a road network graph $G=(V,E)$, and a set $Q=\{(s_{1},d_{1}),(s_{2},d_{2}),...,(s_{q},d_{q})\}$ of trip queries in the form of source-destination pairs, where $|V|=n$, $|E|=m$, $|Q|=q$, $Q.S=\{s_{1}, s_{2}, ..., s_{q}\}$, $Q.D=\{d_{1}, d_{2}, ..., d_{q}\}$, and $\forall i \in [1,q] \quad (s_{i},d_{i} \in V)$. First, we define the problem of finding an optimal pair of end-stops. Second, we formulate the problem of determining the optimal route and the intermediate stops. Last, we suggest several variants of the second task. In the discussion below, let $SPC(u,v)$ denote the shortest path cost to a node $v$ from a node $u$ in the road network graph. For the purpose of presenting each problem, without loss of generality, we may assume that all roads are bi-directional and have the same cost in either direction.

\subsection{The Optimal End-Stops}
\label{poes}
In this problem, the input sources (resp. destinations) are co-located. Our goal is to determine a start-stop $st \in V$, and an end-stop $en \in V$ of a vehicle such that the following aggregate cost function is minimized:
\begin{align*}
C_{1}(st,en)=SPC(st,en)+\sum\limits_{i=1}^{q} (SPC(s_{i},st)+SPC(en,d_{i}))
\end{align*}

\subsection{The Optimal Route and Intermediate Stops}
\label{pors}
In this problem, in addition to $G$, and $Q$, we are given a start-stop $st \in V$, and an end-stop $en \in V$. Our task is to find an ordered sequence of stops $P=[P(1),P(2),...,P(t)]$ of a vehicle, where $\mathbf{len}~P=t$, and $(P(1)=st) \land (P(t)=en) \land (\forall i \in [1,t] \quad (P(i) \in V))$. The vehicle's route is the collection of shortest paths among consecutive stops in $P$. Let us define the functions $f_{s}: \{(Q.S,V)\} \rightarrow \{True,False\}$, and $f_{d}: \{(V,Q.D)\} \rightarrow \{True,False\}$. $f_{s}(s_{i},P(j))$ is $True$ when the $i'th$ passenger starting from $s_{i}$ enters the vehicle at $P(j)$, $False$ otherwise.  Similarly, $f_{d}(P(k),d_{i})$ is $True$ only when the $i'th$ passenger going to $d_{i}$ exits the vehicle at $P(k)$. A passenger gets on (resp. off) at a unique stop, where the entry point precedes the exit point in the sequence $P$. We need to compute $P$ in such a way that minimizes the following aggregate cost function:
\begin{align*}
C_{2}(P)=\sum\limits_{i=1}^{t-1} SPC(P(i),P(i+1))+\sum\limits_{i=1}^{q} \sum\limits_{j=1}^{t} \sum\limits_{k=1}^{t}\\(SPC(s_{i},P(j))*[f_{s}(s_{i},P(j))]+SPC(P(k),d_{i})*[f_{d}(P(k),d_{i})]),\\where \quad \forall i \in[1,q] \quad ((\exists! j \in[1,t] \quad ([f_{s}(s_{i},P(j))]=1))\land\\(\exists! k \in[1,t] \quad ([f_{d}(P(k),d_{i})]=1))\land(j<=k))
\end{align*}
Here, $\exists!$ denotes uniqueness quantification. For any boolean statement $S$, $[S]$ equals $1$ when $S$ is $True$, $0$ otherwise.

Below we recommend several variants under a number of different constraints.

\subsubsection{Constraint on Each User's Lone Path Length Before Entering or After Exiting}
\label{pc1}
In the cost function $C_{2}$, we may add the following constraint to check the maximum length a user may travel before getting on, or after getting off the vehicle:
\begin{align*}
\max_{i=1}^{q} \max_{j=1}^{t} \max_{k=1}^{t}(SPC(s_{i},P(j))*[f_{s}(s_{i},P(j))],\\SPC(P(k),d_{i})*[f_{d}(P(k),d_{i})])<=R_{1},\quad where \quad R_{1} \in [0,\infty)
\end{align*}
The limit $R_{1} \rightarrow \infty$ technically means that there is no restriction on a user's path; while, with $R_{1}=0$, the problem becomes identical to the traveling salesman path problem $(TSPP)$ \cite{DBLP:journals/mp/LamN08}.

\subsubsection{Constraint on the Vehicle's Route Length}
\label{pc2}
The following constraint on $C_{2}$ limits the vehicle's length of travel:
\begin{align*}
\sum\limits_{i=1}^{t-1} SPC(P(i),P(i+1))<=R_{2}*SPC(st,en),\quad where \quad R_{2} \in [1,\infty)
\end{align*}
When $R_{2}=1$, the vehicle travels in its shortest path. Contrarily, when $R_{2} \rightarrow \infty$, the task essentially remains the same as the original problem.

\subsubsection{Constraint on the Entering/Exiting Passenger-Cardinality at a Stop}
\label{pc3}
Requiring a minimum number of passengers to get on-or-off the vehicle at an intermediate stop may be one way to limit the total number of stops. The following constraint imposes this limitation:
\begin{align*}
\max_{i=2}^{t-1} (\sum\limits_{j=1}^{q} ([f_{s}(s_{j},P(i))]+[f_{d}(P(i),d_{j})]))>=R_{3},\\where \quad R_{3} \in [0,2*q]
\end{align*}
At one extreme, $R_{3}=0$, or $R_{3}=1$ is equivalent to there being no constraint. At the other extreme, $R_{3}=2*q$ means that $2<=t<=3$, i.e., there is at most one intermediate stop.

\subsubsection{Constraint on the Total Number of Stops}
\label{pc4}
We may limit the total number of stops by adopting a more straightforward constraint:
\begin{align*}
2<=t<=R_{4}, \quad where \quad R_{4} \in [2,2*q+2]
\end{align*}
$R_{4}=2$ means that the vehicle does not stop midway at all; while $R_{4}=2*q+2$ suggests that the driver can always freely stop to pick up/drop off a single passenger.

\subsubsection{The Weighted Version}
\label{pc5}
Instead of assigning equal weights to the route cost of the vehicle, and the total lone travel cost of the passengers, we may discriminate as follows:
\begin{align*}
C_{2}(P)=R_{5}*(\sum\limits_{i=1}^{t-1} SPC(P(i),P(i+1)))+(1-R_{5})*(\sum\limits_{i=1}^{q} \sum\limits_{j=1}^{t} \sum\limits_{k=1}^{t}\\(SPC(s_{i},P(j))*[f_{s}(s_{i},P(j))]+SPC(P(k),d_{i})*[f_{d}(P(k),d_{i})])),\\where \quad R_{5} \in[0.\overline{3},1] \land \forall i \in[1,q] \quad ((\exists! j \in[1,t] \quad ([f_{s}(s_{i},P(j))]\\=1))\land(\exists! k \in[1,t] \quad ([f_{d}(P(k),d_{i})]=1))\land(j<=k))
\end{align*}
When $R_{5}=0.\overline{3}$, the weighted version reduces to the $TSPP$ problem (see \cite{DBLP:journals/tkde/LiQYM16} for a proof). Contrarily, when $R_{5}=1$, the shortest path between the end-stops is the optimal route of the vehicle.

\bigskip
As discussed in Section~\ref{intro}, each constraint serves a specific purpose. We may easily formulate more variants by imposing more than one of the above four constraints at once.

\subsection{Overview}
\label{ov}
We organize the remainder of our paper as follows. First, in Section~\ref{relworks}, we review the existing literature. Then, in Section~\ref{oes}, we provide algorithms to solve the \emph{optimal end-stops} problem. We demonstrate the straightforward baseline technique in Section~\ref{baseline} and present our more efficient approach in Section~\ref{app}. After that, in Section~\ref{oris}, we investigate the \emph{optimal route and intermediate stops} problem and its variants. We provide the exact solution approach in Section~\ref{opt}, and the heuristic solution in Section~\ref{app2}. Then, in Section~\ref{var}, we discuss the solutions of the variants; in Section~\ref{tspp}, we demonstrate the relation of our second problem to the $TSPP$ \cite{DBLP:journals/mp/LamN08}. Finally, in Section~\ref{exp}, we show the results of our experiments.

We assume that our each algorithm takes as input, a reduced road network graph that makes sense, i.e., beyond which, neither the vehicle nor the passengers require traveling. The reduced graph may be a reasonably large elliptical section of the original graph, or a sub-graph with a region-specific boundary, e.g., the road network graph of the San Francisco city \cite{DBLP:journals/geoinformatica/Brinkhoff02}. We also assume that the path queries input to our problems meet the requirements mentioned in Section~\ref{intro}. We consider city scale graphs, rather than continent scale ones, since they are more practical for the scope of our queries.

\section{Related Works}
\label{relworks}
In this paper, we have introduced two types of problems - the optimal end-stops $(OES)$ query, and the optimal route and intermediate stops $(ORIS)$ query. In the existing literature, our $OES$ and $ORIS$ problems are closely related to the ride-sharing problem in road networks \cite{IDM-LAB:journals/elsevier/Koen13i}. In recent years, several studies, \cite{DBLP:conf/hicss/LiHZ17}, \cite{DBLP:conf/huc/CiciMFL14}, \cite{DBLP:conf/amcis/WangAS17}, \cite{Alonso-Mora17012017}, have demonstrated the benefits of ride-sharing in reducing the traffic congestion \cite{DBLP:conf/hicss/LiHZ17}, \cite{DBLP:conf/huc/CiciMFL14}, the number of DWI fatalities \cite{DBLP:conf/amcis/WangAS17}, and the greenhouse gas emission \cite{Alonso-Mora17012017}. \cite{DBLP:conf/percom/Stach11} shows how a ride-sharing system may save time, money and the environment. Our techniques complement the existing ride-sharing approaches, \cite{IDM-LAB:journals/elsevier/Koen13i}, by computing the optimal route and stops of a vehicle for a group of assigned passengers.

The challenges in the ride-sharing system come from two directions. First, to dynamically match the passengers, requesting shared rides, to appropriate vehicles. Second, to compute the best route of each vehicle and its pick-up and drop-off locations for the passengers assigned to it. Neither task is trivial. Several works in the existing literature address the first problem \cite{DBLP:journals/pvldb/HuangBJW14}, \cite{DBLP:journals/tkde/MaZW15}, \cite{DBLP:conf/icde/MaZW13}. \cite{DBLP:journals/pvldb/HuangBJW14} presents an efficient algorithm based on the kinetic tree, which finds the appropriate assignment with a service guarantee. In \cite{DBLP:conf/icde/MaZW13}, a system, named T-share, performs dynamic vehicle-passenger matching for the purpose of Taxi ride-sharing. \cite{DBLP:journals/tkde/MaZW15} introduces a Spatio-temporal index structure, which facilitates taxi searching under a set of constraints such as the time-window constraints, and the monetary constraints. These algorithms concentrate on efficiently assigning the passengers to the vehicles in the system in real-time. In this paper, we mainly focus on overcoming the second challenge of determining the optimal route and stops of a vehicle through solving the $OES$ and the $ORIS$ problems.

A general class of NP-hard problems, namely, the vehicle routing problem $(VRP)$ \cite{DBLP:journals/corr/Munari16}, \cite{DBLP:journals/candie/BraekersRN16}, \cite{DBLP:journals/corr/MahmoudiZ15}, \cite{DBLP:journals/dam/TothV02}, is somewhat related to both the $OES$ and the $ORIS$ problems. To detail the myriad problems under the heading of $VRP$ is beyond the scope of our discussion. The general objective of these studies is to minimize the global cost of delivery of commodities or passengers, meeting a myriad of constraints, by using a fleet of vehicles. Remember that the goal of our problems is to reduce the total cost of travel by letting the passengers share a single vehicle; each passenger journeys alone before entering and after leaving the vehicle. Of course, sharing a common large vehicle in a highway helps in reducing the global cost to some extent. However, subsets of the passengers also share other common paths in the suburbs. If, whenever more than one passenger shared a path in common, they had traveled collectively, the global travel cost would be the minimum; the $VRP$ problem aims to achieve this objective, usually for the delivery of goods. Another related problem, with a similar goal, is the collective travel planning $(CTP)$ query \cite{DBLP:journals/tkde/ShangCWJWK16}. The $CTP$ problem aims at finding the lowest cost route connecting multiple sources and a destination, via at most $k$ meeting points from a set of prospective locations. The motivation behind the $VRP$ and the $CTP$ queries are to achieve an environment-and-cost-friendly transportation that reduces traffic congestion, energy consumption, and greenhouse gas emission. Mathematically, these are similar to the Steiner diagram problems \cite{DBLP:journals/jda/BlasumHOW07}, \cite{DBLP:reference/algo/Chuzhoy08}, \cite{DBLP:journals/networks/DreyfusW71}, which would find an optimal acyclic sub-graph of the road network, connecting the query nodes of our problems. In general, all these problems have integer programming, or integer network flow formulations; a myriad of approximations and heuristics are available to cope with practical scenarios. However, our problems are fundamentally different from these in that we expect the passengers to travel collectively, only in the most major shared route, using a single vehicle. Again, unlike the $VRP$ and the $CTP$ problems, our queries do not require that the vehicle passes through any node(s) specified in the input. Therefore, the techniques in \cite{DBLP:journals/corr/Munari16}, \cite{DBLP:journals/candie/BraekersRN16}, \cite{DBLP:journals/corr/MahmoudiZ15}, \cite{DBLP:journals/dam/TothV02}, \cite{DBLP:journals/tkde/ShangCWJWK16}, \cite{DBLP:journals/jda/BlasumHOW07}, \cite{DBLP:reference/algo/Chuzhoy08}, and \cite{DBLP:journals/networks/DreyfusW71} do not help in solving our problems. The $VRP$ and the $CTP$ queries fare better in addressing the first challenge of the ride-sharing system than in dealing with the second one. A ride-sharing system may utilize these queries to assign the passengers to different vehicles - possibly recommending each passenger to travel by using more than one vehicle en-route. Once the assignments are complete, our $OES$ and $ORIS$ queries may help in finding the optimal route for each vehicle.

In the $OES$ problem, given a cluster of co-located sources, and another of co-located destinations, we are to find an optimal pair of end-stops for a vehicle, which will carry the passengers from the source cluster to the destination cluster. Users demanding a vehicle, using a location-based service, do not automatically form clusters. The initial task is to group them in a way that satisfies our input requirements. In this paper, we do not provide an algorithm for grouping passengers. Instead, we assume that an existing clustering algorithm such as \cite{DBLP:conf/ssd/MahmudAAHN13}, which partitions the queries into batches, has already performed the grouping. \cite{DBLP:conf/ssd/MahmudAAHN13} divides the path queries into groups, where each group comprises the queries from a source cluster to a destination cluster. We take each output query group of \cite{DBLP:conf/ssd/MahmudAAHN13} as input to our algorithm and focus on determining the optimal end-stoppages for the corresponding vehicle.

To the best of our knowledge, our $OES$ problem is new in the literature. A related problem is the optimal meeting point $(OMP)$ query in road networks \cite{DBLP:journals/pvldb/YanZN11}, \cite{DBLP:journals/kais/YanZN15}. In the $OMP$, the query is a set of nodes; the target is to determine a meeting location such that the aggregate cost of travel from the query nodes to that location is the minimum. The $OMP$ query is fundamentally different from our $OES$ query. In the former, the objective cost is a function of the distances of the query nodes to the meeting point. Contrarily, in the latter, the cost function depends on both the individual path costs of the users and the route cost of the vehicle. If we provide the co-located sources (resp. destinations) of our problem as input to the $OMP$ query \cite{DBLP:journals/pvldb/YanZN11}, it may find an approximation to our optimal start-stop $st$ (resp. end-stop $en$). However, since the $OMP$ query does not take into account the location and orientation of the other distant cluster, it will fail to compute a reasonable answer to our query, in most instances. Therefore, we cannot use the existing solutions for the $OMP$ query to solve the $OES$ problem.

Other problems, related to, yet, very different from the $OES$, are the group nearest neighbor $(GNN)$ \cite{DBLP:conf/icde/PapadiasSTM04}, the group k-nearest neighbors $(GKNN)$ \cite{DBLP:journals/jgs/Safar08}, the k-optimal meeting points $(k-OMP)$ \cite{DBLP:conf/dnis/TiwariK13a}, and the optimal location $(OL)$ \cite{DBLP:reference/gis/Xiao17} queries in the road network. Unlike the $OES$, and like the $OMP$, none of these queries take a pair of clusters (the source cluster and the destination cluster) as input. Thus, the techniques in \cite{DBLP:conf/icde/PapadiasSTM04}, \cite{DBLP:journals/jgs/Safar08}, \cite{DBLP:conf/dnis/TiwariK13a}, and \cite{DBLP:reference/gis/Xiao17} do not apply in the context of the $OES$ problem.

Our $OES$ query is different from the point-to-point shortest path query in two aspects. First, it takes multiple source-destination pairs as input. Second, it aims to optimize a different cost function, which is a summation of the vehicle's path cost and the passenger's solo travel costs. Therefore, the traditional shortest path algorithms such as the $Dijkstra$ \cite{DBLP:books/daglib/0023376}, \cite{DBLP:conf/anziis/EklundKP96}, and the $A^{*}$ \cite{DBLP:books/lib/RussellN03} cannot answer our query. Similarly, we cannot use faster hierarchy based approaches like \cite{DBLP:conf/ssd/ShekharFG97}, \cite{DBLP:conf/wea/GeisbergerSSD08}, and \cite{DBLP:conf/sigmod/ZhuMXLTZ13} either. However, we benefit from the $Dijkstra$'s algorithm \cite{DBLP:books/daglib/0023376}, \cite{DBLP:conf/anziis/EklundKP96}, the bidirectional search \cite{DBLP:conf/aaai/DavisPS84}, and the group shortest path approach \cite{DBLP:conf/mdm/RezaAH15}. \cite{DBLP:conf/mdm/RezaAH15} introduces a technique to process a batch of shortest path queries simultaneously, based on the path-coherence property of road networks. Although \cite{DBLP:conf/mdm/RezaAH15} cannot solve our problem, we borrow the intuition behind its simultaneous search to develop an efficient algorithm in Section~\ref{app} that answers the $OES$ query.

In the $ORIS$ problem, given the end-stoppages and a set of path queries from a group of users, we are to compute an optimal route for a vehicle, as a sequence of intermediate stops. Usually, in a ride-sharing system, there is a fleet of vehicles for providing the passengers with shared rides. The first task is to divide the passengers into groups and assign each group to a vehicle; we assume that a vehicle-passenger matching algorithm, e.g., \cite{DBLP:journals/pvldb/HuangBJW14}, \cite{DBLP:journals/tkde/MaZW15}, \cite{DBLP:conf/icde/MaZW13}, and \cite{DBLP:conf/gis/AlarabiCZMB16}, has already done that. In this paper, we center upon the task of determining the optimal route for a single vehicle serving a group of passengers.

In Section~\ref{tspp}, we demonstrate the $ORIS$ query's relation to the well-known NP-hard traveling salesman path problem $(TSPP)$ \cite{DBLP:journals/mp/LamN08}. We show that the $TSPP$ is a fixed-parameter version of the $ORIS$ problem. In particular, we prove that any algorithm, which solves a variant of our problem for any arbitrary value of a parameter, also solves the $TSPP$. Be that as it may, we cannot use the solution techniques of \cite{DBLP:journals/mp/LamN08} to answer the $ORIS$ query. Our $ORIS$ query is also closely related to the trip planning $(TP)$ problem \cite{DBLP:conf/ssd/LiCHKT05}, \cite{DBLP:conf/ssd/HashemHAK13}, and the optimal sequenced route $(OSR)$ problem \cite{DBLP:journals/vldb/SharifzadehKS08}, \cite{DBLP:conf/gis/CostaNMTPM15}, \cite{DBLP:conf/mdm/SamroseHBAUM15}, \cite{DBLP:journals/geoinformatica/ChenKSZ11}, \cite{DBLP:journals/geoinformatica/SharifzadehS08}. The $TP$ query aims at finding the optimal trip schedule that starts from a source location, goes through several points of interests $(POI)$ such as restaurants, theaters, zoos, etc., and ends at a destination node; the $TP$ calculates a schedule for either a single user \cite{DBLP:conf/ssd/LiCHKT05}, or a group \cite{DBLP:conf/ssd/HashemHAK13}. Similarly, the $OSR$ problem, introduced in \cite{DBLP:journals/vldb/SharifzadehKS08}, is to determine a minimum-cost route from a source to a destination, passing through several typed $POI$s in a sequence, specified on the types, in the road network. Our $ORIS$ query is different from both the $TP$ and the $OSR$ queries in the following aspects. First, unlike in the other two problems, in the $ORIS$ query, the optimal route of the vehicle does not necessarily pass through a set of $POI$s. Indeed, in our problem, most passengers may travel some distance individually before getting on (resp. after getting off) the vehicle. The pick-up and the drop-off locations, through which our vehicle travels, are not query points; they are the output of our algorithms. Second, in the $ORIS$ problem, the nodes do not have a type information. The query nodes are simply the source and the target locations of the passengers. Last, our $ORIS$ query does not require to meet any sequence constraint; rather, it is the objective of our algorithms to report the intermediate stops in an optimal sequence along the vehicle's optimal route. As a result of these differences, the solution approaches to \cite{DBLP:conf/ssd/LiCHKT05}, \cite{DBLP:conf/ssd/HashemHAK13}, \cite{DBLP:journals/vldb/SharifzadehKS08}, \cite{DBLP:conf/gis/CostaNMTPM15}, \cite{DBLP:conf/mdm/SamroseHBAUM15}, \cite{DBLP:journals/geoinformatica/ChenKSZ11}, and \cite{DBLP:journals/geoinformatica/SharifzadehS08} are not applicable in solving the $ORIS$ problem. Another query, which is somewhat related to ours, is the keyword-aware optimal route $(KOR)$ problem \cite{DBLP:conf/icde/CaoCCGPX13}. The $KOR$ query searches for the optimal route that goes through a set of nodes covering an input set of keywords. This problem is also dramatically different from our $ORIS$ query; hence, we cannot apply the techniques presented in \cite{DBLP:conf/icde/CaoCCGPX13} either.

Several ride-sharing path-planning approaches, such as \cite{DBLP:conf/socs/DrewsL13}, and \cite{DBLP:conf/atmos/GeisbergerLNSV10}, also relates to our $ORIS$ query. However, these require that the optimal detour route, from the source $s$ to the destination $t$, includes a sub-route between two nodes - $s'$ and $t'$ - specified in the query. Hence, the route returned by these ride-sharing queries obligatorily passes through two query nodes. Contrarily, in our $ORIS$ problem, the vehicle's route does not need to include any query node other than the end-stoppages. This difference renders us unable to use the algorithms in \cite{DBLP:conf/socs/DrewsL13} and \cite{DBLP:conf/atmos/GeisbergerLNSV10} to answer our query.

Another interesting approach, related to our $ORIS$, is the carpooling problem \cite{DBLP:conf/kdd/GeLXC11}, \cite{DBLP:conf/sensys/ZhangLZLLH13}, \cite{DBLP:journals/tetc/ZhangHLLS14}. The $coRide$ system, presented in \cite{DBLP:conf/sensys/ZhangLZLLH13}, comprises a mobile app for passenger clients, onboard hardware devices in each taxi dedicated for notifications and data gathering, and a cloud dispatching server. The cloud server performs the vehicle-passenger matching, recommends routes, and estimates fare of each passenger. \cite{DBLP:conf/sensys/ZhangLZLLH13} proposes a win-win fare model that benefits both the passengers and the drivers economically. \cite{DBLP:conf/sensys/ZhangLZLLH13} tackles the passenger assignment and the route recommendation problems simultaneously; it provides an integer programming formulation with constraints such as the number of available taxicabs, the capacity of each taxicab, and time-window requirement of each passenger. Besides an exponential-time optimal solution, \cite{DBLP:conf/sensys/ZhangLZLLH13} proposes a fast 2-approximation algorithm for practical purposes. The route calculation in our $ORIS$ query is significantly different from that in the $coRide$ problem as follows. First, in the $coRide$, a vehicle picks a passenger up (resp. drops him/her off) at his/her query node. Contrarily, in the $ORIS$, the vehicle does not necessarily travel through each query node. Second, the $coRide$ system requires that some vehicle, from among the fleet, picks a particular passenger up within a specified time window. This constraint limits the matching possibilities and the order in which a vehicle may pick its passengers up. Our $ORIS$ does not handle any time-window constraint; thus, it has more choices for its route. Due to these factors, even a $coRide$ system with a single vehicle is different from the $ORIS$ query. Therefore, we cannot use the technique in \cite{DBLP:conf/sensys/ZhangLZLLH13} to solve the $ORIS$.

Like our $ORIS$, the optimal multi-meeting-point route $(OMMPR)$ search problem also tackles the second challenge of the ride-sharing, namely, determining a vehicle's route for a group of matched passengers \cite{DBLP:journals/tkde/LiQYM16}. However, \cite{DBLP:journals/tkde/LiQYM16} solves only the fifth variant of our problem (Section~\ref{pc5}). It provides two straightforward dynamic programming solutions - the $basic$ and the $grow$ methods, and two optimized ones - the $bidirectional$ and the $bidirectional-bounded$ techniques; all four algorithms compute the exact answers. However, even the most optimized of these four approaches, namely, the $bidirectional-bounded$ method, has both time and space complexities exponential in the number of query nodes. Therefore, the techniques presented in \cite{DBLP:journals/tkde/LiQYM16} do not scale well for a reasonably large number of users. In other words, the algorithms in \cite{DBLP:journals/tkde/LiQYM16} are only suitable for a small 5/6-seated taxi-ride-sharing. Contrarily, the heuristic technique, which we have developed in Section~\ref{app2}, returns an answer instantly for even a large number of queries, like fifty or more queries. Thus, our approach is more suitable for ride-sharing using a larger vehicle such as a mini-bus, or a bus. Our experiments, in Section~\ref{exp}, illustrates that the gain in execution time and memory is worth the little loss of accuracy by our technique. To the best of our knowledge, our heuristic algorithm in Section~\ref{app2} is the first of its kind, which efficiently tackles the second challenge of ride-sharing for a large number of passengers per vehicle. Another contribution of our paper is that we have formulated some useful variants in Section~\ref{pors} and proposed their solutions in Section~\ref{var}.

Notice that the shortest path from the start-stop $st$ to the end-stop $en$ is obviously not the answer to the $ORIS$ query. Therefore, we cannot directly apply the shortest path computation algorithms, e.g., \cite{DBLP:books/daglib/0023376}, \cite{DBLP:books/lib/RussellN03}, \cite{DBLP:conf/ssd/ShekharFG97}, \cite{DBLP:conf/wea/GeisbergerSSD08}, and \cite{DBLP:conf/sigmod/ZhuMXLTZ13}, to solve our problem. Be that as it may, we develop our exponential-time exact method in Section~\ref{opt} by modifying the $Dijkstra$'s algorithm \cite{DBLP:books/daglib/0023376}; we define each sub-problem state as a node-set-of-queries pair. We also base the intuition for our near-optimal polynomial-time heuristic search procedure, presented in Section~\ref{app2}, on the greedy strategies of the $Dijkstra$'s \cite{DBLP:books/daglib/0023376}, \cite{DBLP:conf/anziis/EklundKP96}, and the $Bellman$-$Ford$-$Moore$'s \cite{DBLP:journals/tcs/JuknaS16} single-source shortest path algorithms.

\section{The Optimal End-Stops}
\label{oes}
In this section, we analyze the problem of finding the optimal end-stops of a vehicle, given path queries from a group of users from co-located sources to co-located destinations. First, we outline the straightforward slow solution. Last, we propose our fast and exact solution.

\subsection{Baseline Solution}
\label{baseline}
The baseline algorithm is pretty straightforward. Initially, we compute the shortest path costs from the query sources to all nodes in the reduced road network graph (as discussed in Section~\ref{ov}), and from all nodes to the query destinations. At each node, we store the sum of the shortest path costs from the sources and the sum of the shortest path costs to the destinations. Then, from each node of the graph, we compute the shortest path costs to all nodes. We consider each pair of nodes as a candidate solution whose cost is determined by the cost function $C_{1}$. Among all candidate solutions, we choose a pair of nodes that minimizes $C_{1}$, as the optimal end points. To compute single-source shortest paths, we use the $Dijkstra$'s algorithm with $Fibonacci\ Heap$ \cite{DBLP:books/daglib/0023376}, which suffices for the purpose of comparing with our proposed solution.

\begin{algorithm}[!htb]
\caption{\lrg{B}$ASELINE$-\lrg{E}$ND$-\lrg{S}$TOPS\ (G, w, Q)$}
\label{algo1}
\begin{small}

\KwIn{A graph $G$, a cost function for edges $w$, and a set of queries $Q$}
\KwOut{An optimal pair of end-points $(st,en)$}

$Compute\ G^{T}$, and $w^{T}$

\For{each query $(s_{i},d_{i}) \in Q$}
{
	$SPC(s_{i},G.V) \leftarrow $ \lrg{D}$IJKSTRA\ (G, w, s_{i})$
	
	$SPC^{T}(d_{i},G.V) \leftarrow $ \lrg{D}$IJKSTRA\ (G^{T}, w^{T}, d_{i})$	
}

\For{each node $v \in G.V$}
{
	$S_{v} \leftarrow \sum\limits_{i=1}^{q} SPC(s_{i}, v)$
	
	$D_{v} \leftarrow \sum\limits_{i=1}^{q} SPC^{T}(d_{i},v)$
}

$opt \leftarrow \infty$, $(st,en) \leftarrow (NIL,NIL)$

\For{each node $u \in G.V$}
{
	$SPC(u,G.V) \leftarrow $ \lrg{D}$IJKSTRA\ (G, w, u)$
	
	\For{each node $v \in G.V$}
	{
		\If{$SPC(u,v)+S_{u}+D_{v} < opt$}
		{
			$opt \leftarrow SPC(u,v)+S_{u}+D_{v}$
			
			$(st,en) \leftarrow (u,v)$
		}
	}
}

\KwRet $(st,en)$

\end{small}
\end{algorithm}

Algorithm~\ref{algo1} illustrates the pseudo-code for the baseline approach. Line 1 computes the transpose graph of $G$, namely $G^{T}=(V,E^{T})$, where $E^{T}=\{(u,v):(v,u) \in E\}$, i.e., $E^{T}$ consists of the edges of $G$ with their directions reversed; it also computes the transpose of the cost function $w$, namely $w^{T}$, which stores the cost of the edges in $E^{T}$. Lines 2-4 calculate the shortest path costs to all nodes from the query nodes. From each source node (resp. destination node), the $DIJKSTRA$ routine computes the single-source shortest paths on the graph $G$ (resp. $G^{T}$). Lines 5-7 calculate the summation of the shortest path costs $S_{v}$ (resp. $D_{v}$) to each node $v \in V$ (resp. the destinations) from the sources (resp. each node $v \in V$). Line 8 initializes the potential optimal value $opt$ for the cost function $C_{1}$ to $\infty$, and the answers $(st,en)$ to $(NIL,NIL)$. The \textbf{for} loop in lines 9-14 considers each node $u \in V$ as a potential start-stop and calculates single-source shortest paths from it in line 10. Then, the \textbf{for} loop in lines 11-14 regards each node $v \in V$ as a potential end-stop, computes $C_{1}(u,v)$ as a candidate solution cost, and finds the optimal solution from among all candidate solutions. Finally, line 15 returns the optimal answers $(st,en)$.

\subsubsection{Complexity Analysis}
\label{baseline_complexity}
In Algorithm~\ref{algo1}, line 1 needs $O(n+m)$ to compute $G^{T}$. Computing single-source shortest paths using the $Dijkstra$'s algorithm with $Fibonacci\ Heap$ takes $O(n \lg n+m)$. Thus, lines 2-4 require $O(q(n \lg n+m))$; usually, $q << n$. To calculate the summations in lines 5-7, $O(nq)$ is needed. The initialization in line 8 is in $O(1)$. In each iteration of the \textbf{for} loop in lines 9-14, line 10 takes $O(n \lg n+m)$ and lines 11-14 demand $O(n)$. Hence, lines 9-14 involve $O(n^2 \lg n+nm)$ computation which is the dominating term in this complexity analysis. In a dense graph, $m=cn^2$, $c$ being a constant; however, in sparse road network graphs, $m=cn$. Therefore, the overall complexity of the baseline solution is $O(n^2 \lg n)$.

We know that an implementation of the $Dijkstra$'s algorithm with $Fibonacci\ Heap$ and Adjacency List representation of the graph requires $O(n+m)$ space. In a sound implementation of Algorithm~\ref{algo1}, besides the space required by the $Dijkstra$'s algorithm, we need only the following space - the summation of costs to each node from the sources, and the summation of costs to the destinations from each node. Thus, additional space needed is $O(n)$; overall space complexity is $O(n+m)$.

\subsubsection{Discussion}
\label{dis}
We may improve the baseline algorithm by replacing each $Dijkstra$-based shortest path computation with a faster approach, e.g., a method based on the contraction hierarchies \cite{DBLP:conf/wea/GeisbergerSSD08}, and another using the arterial hierarchies \cite{DBLP:conf/sigmod/ZhuMXLTZ13}. However, no matter which method we use to compute the shortest paths, we still need to regard each pair of nodes in the graph as possible end-stoppages. Hence, calculating the all pair shortest paths is a must in the baseline technique. Thus, even using the state-of-the-art in the shortest path computation \cite{DBLP:conf/sigmod/ZhuMXLTZ13}, which answers a query in constant time, cannot improve the complexity of the brute-force technique beyond $O(n^2)$. Besides, although \cite{DBLP:conf/sigmod/ZhuMXLTZ13} has a constant time complexity per point-to-point path query, it depends on massive pre-computation. As we show shortly, our proposed solution works many times faster than the $O(n^2)$ bound. Therefore, we judge that implementing \cite{DBLP:conf/wea/GeisbergerSSD08} or \cite{DBLP:conf/sigmod/ZhuMXLTZ13} is not worth a mere $O(\lg n)$-factor gain in the complexity of the baseline method, which we only use for comparing with our much faster solution approach.

\subsection{Fast Solution}
\label{app}
In this section, we present a novel algorithm to compute the optimal end-stoppages. The baseline approach computes path cost between each pair of nodes in the reduced graph. It does not exploit the fact that the query sources (resp. destinations) are co-located or any property of the underlying road network. Hence, the brute-force solution is inefficient. In our new algorithm, we develop a search technique that utilizes the path coherence property of road networks. Our approach performs a simultaneous search from all the sources and another from all the destinations. In this method, the existence of shared routes among queries helps reduce the expected execution time. In practice, our algorithm is an order of magnitude faster than the baseline technique.

As mentioned above, we perform a search involving the query sources and another including the query destinations. In the former, we compute the minimum total cost of travel to each node $v \in V$ in the reduced road network (as discussed in Section~\ref{ov}), when the passengers travel alone to some node, say $st$, and the vehicle carries them from $st$ to $v$. In the latter, we calculate a similar minimum traveling cost from each node $v \in V$, when the vehicle conveys the passengers from $v$ to some node, say $en$, and each passenger journeys individually from $en$ to his/her respective destination.

Since both searches are similar, we detail only the search from the query sources. Throughout the search procedure, we maintain a frontier using a priority queue $PQ$. We define the search frontier as a set of nodes, from each of which, we are yet to branch some queries to its adjacent nodes. Initially, $PQ$ comprises each query source node with the respective query waiting to be branched from that node. We relatively order each node in the search frontier $PQ$ by a cost associated with that node; we term this cost as the node's key cost. We shall outline the computation of a node's key cost shortly. Each time, we extend the search space by picking a node from $PQ$ with the minimum key cost and branch the queries waiting at that node to its adjacent nodes. It is easier to visualize our search technique as individual searches from the query nodes, running in parallel threads. The difference is that ours is single search, where more than one query may wait at a node simultaneously; when a node's turn arrives, we branch the queries waiting at it to each neighbor at once. We term the process of branching queries to an adjacent node as relaxation. Notice, at any moment during our search, each query likely reaches a different subset of nodes in the reduced road network graph. Hence, at a particular time, for each node in the graph, we can find a subset of queries, whose individual search space has reached that node; we term this subset as the node's currently-reaching-queries. We maintain and update the following at each node:

\begin{itemize}
\item The node's key cost, which we use for relatively ordering it in the search frontier $PQ$.
\item The node's parent, which is its predecessor on the vehicle's route; the parent is $NIL$, when the node is not on the vehicle's path or is the start-stop.
\item Each currently-reaching-query and the shortest path from the corresponding source node; we let this information persist even after we relax the queries waiting at it.
\end{itemize}

Let us delineate the relaxation of the queries waiting at a node $u$, to a node $v$. First, for each query $q_{i}$ waiting at $u$, we determine the shortest path cost from its source node $s_{i}$ to $v$ by considering the path through $u$ as an option. We may need to update the costs of some already-reaching-queries if the paths through $u$ turn out a better option. Again, we may require inserting some new queries at $v$, if they reach $v$ for the first time during this relaxation. Second, we compute the key cost of $v$ as the minimum of the below three options:

\begin{enumerate}
\item Summation of the costs of the queries currently reaching $v$, in case we have updated/inserted some query cost at $v$ through $u$.
\item Key cost of $u$ + edge cost between $u$ and $v$, in case all the queries were waiting at $u$.
\item Previous key cost of $v$, in case no new query has reached $v$ through $u$.
\end{enumerate}

$Option\ 1$ represents the situation when each passenger is still traveling alone; if all the queries are currently reaching $v$, then it means that they have just boarded the bus. Hence, we set the parent of $v$ to $NIL$ if $option\ 1$ is the minimum. $Option\ 2$ indicates that the vehicle has carried all the passengers from $u$ to $v$; we, therefore, count the edge cost between $u$ and $v$ only once as part of the vehicle's routing cost, not for the individual passengers, since they are already onboard. If this option is the minimum, we set the parent of $v$ to $u$. $Option\ 3$ simply tells that no new query has reached $v$ and that its previous key cost is better than the other options; thus, we keep its key cost and parent unchanged. Notice that this option does not necessarily mean that we have not performed any update during the relaxation. Indeed, we may have updated some query cost at $v$ through $u$; however, the former key cost may still be the minimum due to a better route of the vehicle through some other node. Last, in the scenario that we have performed any update in the first or the second step, we consider $v$ as a frontier node in the search space. Even when $option\ 3$ is the minimum, we push $v$ to the frontier $PQ$ as long as we have updated at least one query cost at $v$, since propagating this update may help obtain better key costs in future relaxations. Recall, we expand the search space by each time picking the minimum-cost-node from $PQ$ and relaxing to its neighbors. We execute the search exhaustively until $PQ$ becomes empty.

\begin{figure}[!tb]
\centering
\includegraphics[width=.90\columnwidth]{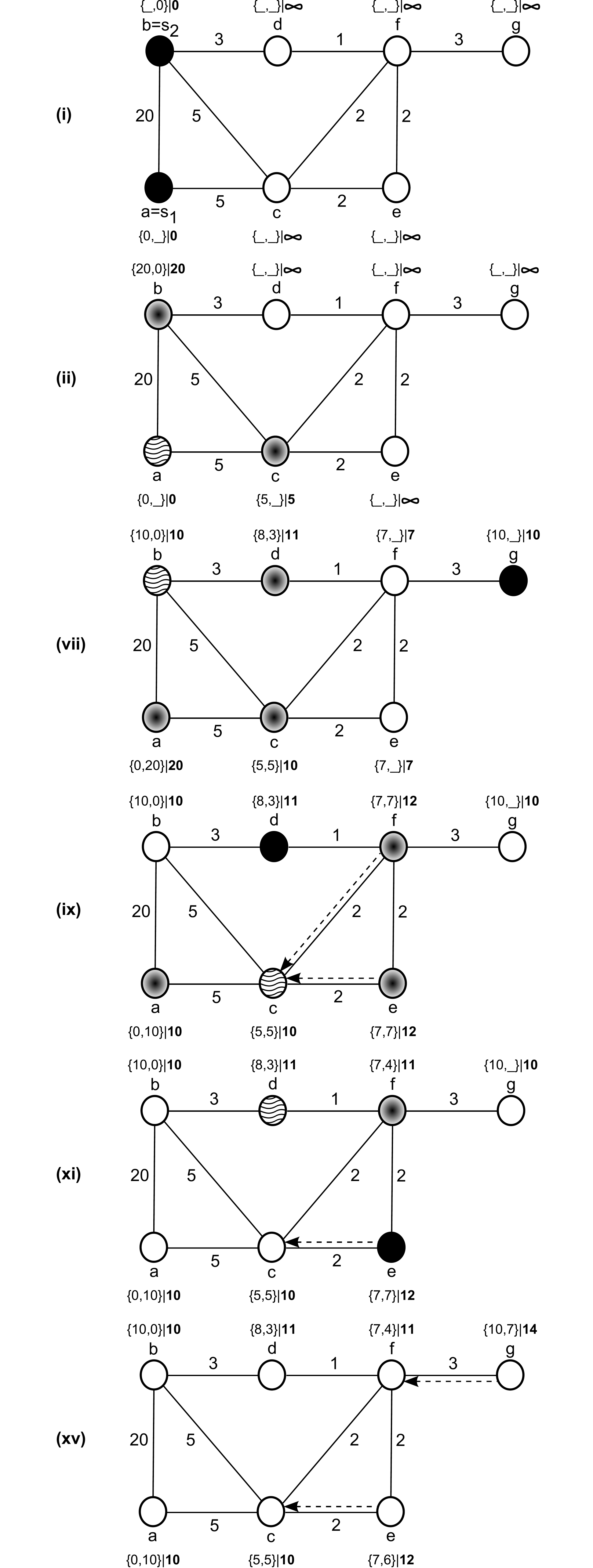}
\caption{Example Relaxations of Algorithm~\ref{algo2} - \\Steps (i), (ii), (vii), (ix), (xi), and (xv).}
\label{fig_algo2}
\end{figure}

We illustrate our search technique with an example in Figure~\ref{fig_algo2}. In this figure, we show several relaxations by our algorithm in a sample graph with synthetic queries. Nodes $a$ and $b$ are the query sources $s_{1}$ and $s_{2}$. In the initialization phase, step $(i)$, we set the key cost of node $a$ (resp. node $b$) to $0$, with query $1$ (resp. query $2$) waiting at that node. For the remainder nodes, we set their key costs to $\infty$, with no waiting queries. We set all parents to $NIL$. Nodes $a$ and $b$ are in the search frontier $PQ$. In step $(ii)$, we remove $a$ from $PQ$ and relax query $1$ from $a$ to $b$, and $c$. After these relaxations, the key cost of $b$ becomes $20$, with both queries waiting, and the key cost of $c$ becomes $5$, with only query $1$ waiting; $option\ 1$ prevails in each case. In the figure, for each node, we show its individual query costs and key cost, e.g., $\{20,0\}|20$ for node $b$. In step $(ii)$, we push $c$ to $PQ$ and update the position of $b$. In the next four steps that we do not illustrate for brevity, we relax from $c$, $e$, $f$, and $d$ respectively. In step $(vii)$, we relax from $b$ to $a$, $c$, and $d$; again, $option\ 1$ dominates in each case. In the omitted step $(viii)$, we relax from $g$. In step $(ix)$, we relax from $c$ to its neighbors. $Option\ 1$ reigns at $a$, $option\ 3$ at $b$, and $option\ 2$ at $e$, and $f$. Node $c$ becomes the parent of $e$, and $f$ each. We push $e$, and $f$ to $PQ$ and update the status of $a$. As we have performed no updates at $b$, we keep its frontier status unchanged. Then, after relaxing from $a$ in step $(x)$, in step $(xi)$, we relax from $d$ to its adjacent nodes. Node $b$ remains unchanged. $Option\ 1$ prevails at node $f$, and we reset its parent back to $NIL$. In step $(xii)$ that we leave out, upon relaxations from $f$, $f$ becomes $g$'s parent, and we update the status of $e$, even when $option\ 3$ dominates. After two more steps, step $(xv)$ illustrates the final standing of each node. For example, node $g$ has the cost values $\{10,7\}|14$, and its parent is node $f$. Apparently, passenger $1$ (resp. passenger $2$) travels on the path $a-c-f$ (resp. $b-d-f$) to get on the vehicle at $f$; then, the vehicle carries them to $g$.

Notice that we may carry out the search from the query destinations easily by running a similar search as above in the transpose of the road network graph. After both searches terminate, for each node $v \in V$ in the road network, we calculate the summation of the two key costs at $v$ computed by the searches as a candidate solution cost. The minimum among all such candidates is our answer cost. Finally, we utilize the parent information, stored during the searches, to determine the optimal end-stops, which we return as the outcome of our algorithm.

\begin{algorithm}[!htb]
\caption{\lrg{F}$AST$-\lrg{E}$ND$-\lrg{S}$TOPS\ (G, w, Q)$}
\label{algo2}
\begin{small}
\setcounter{AlgoLine}{0}

\KwIn{A graph $G$, a cost function for edges $w$, and a set of queries $Q$}
\KwOut{An optimal pair of end-points $(st,en)$}

$Compute\ G^{T}$, and $w^{T}$

$(d_{s},\pi_{s},T_{s}) \leftarrow $ \lrg{G}$ROUP$-\lrg{Q}$UERY$-\lrg{S}$EARCH\ (G, w, Q.S)$

$(d_{d},\pi_{d},T_{d}) \leftarrow $ \lrg{G}$ROUP$-\lrg{Q}$UERY$-\lrg{S}$EARCH\ (G^{T}, w^{T}, Q.D)$

$opt \leftarrow \infty$, $mid \leftarrow NIL$

\For{each node $v \in G.V$}
{
	\If{$(|T_{s}(v)|=q) \land (|T_{d}(v)|=q) \land (d_{s}(v)+d_{d}(v)<opt)$}
	{
		$opt \leftarrow d_{s}(v)+d_{d}(v)$
		
		$mid \leftarrow v$
	}
}

$(st,en) \leftarrow $ \lrg{C}$OMPUTE$-\lrg{E}$ND$-\lrg{S}$TOPS\ (mid, \pi_{s}, \pi_{d})$

\KwRet $(st,en)$

\end{small}
\end{algorithm}

\begin{algorithm}[!htb]
\caption*{\lrg{F}$AST$-\lrg{E}$ND$-\lrg{S}$TOPS\ (G, w, Q)$ \textbf{cont.}}
\label{algo21}
\func{\lrg{G}$ROUP$-\lrg{Q}$UERY$-\lrg{S}$EARCH\ $}{$G, w, N$}{
\begin{small}

$(d, \pi, T) \leftarrow $ \lrg{I}$NIT$-\lrg{G}$ROUP$-\lrg{Q}$UERIES\ (G, N)$

$PQ \leftarrow G.V$

\While{$PQ \neq \emptyset$}
{
	$u \leftarrow $ \lrg{E}$XTRACT$-\lrg{M}$IN\ (PQ)$
	
	\For{each node $v \in G.Adj[u]$}
	{
		$(upd,d,\pi,T) \leftarrow $ \lrg{M}$ERGE\ (u,v,w(u,v),d,\pi,T)$
		
		\If{$(|T(u)|=q) \land (d(u)+w(u,v)<d(v))$}
		{
			$upd \leftarrow True$
			
			$d(v) \leftarrow d(u)+w(u,v)$
			
			$\pi(v) \leftarrow u$
		}
		
		\If{$upd=True$}
		{
			\If{\large{E}\small$XISTS\ (PQ, v)$}
			{
				\lrg{U}$PDATE$-$KEY\ (PQ, v, d(v))$
			}
			\Else
			{
				\lrg{I}$NSERT\ (PQ, v)$
			}
		}
	}
}

\KwRet $(d, \pi, T)$

\end{small}}
\end{algorithm}

\begin{algorithm}[!htb]
\caption*{\lrg{F}$AST$-\lrg{E}$ND$-\lrg{S}$TOPS\ (G, w, Q)$ \textbf{cont.}}
\label{algo22}
\func{\lrg{C}$OMPUTE$-\lrg{E}$ND$-\lrg{S}$TOPS\ $}{$mid, \pi_{s}, \pi_{d}$}{
\begin{small}

$(st,en) \leftarrow (mid,mid)$

\While{$\pi_{s}(st) \neq NIL$}
{
	$st \leftarrow \pi_{s}(st)$
}

\While{$\pi_{d}(en) \neq NIL$}
{
	$en \leftarrow \pi_{d}(en)$
}

\KwRet $(st,en)$

\end{small}}
\end{algorithm}

\begin{algorithm}[!htb]
\caption*{\lrg{F}$AST$-\lrg{E}$ND$-\lrg{S}$TOPS\ (G, w, Q)$ \textbf{cont.}}
\label{algo211}
\func{\lrg{I}$NIT$-\lrg{G}$ROUP$-\lrg{Q}$UERIES\ $}{$G, N$}{
\begin{small}

\For{each node $v \in G.V$}
{
	$d(v) \leftarrow \infty$

	$\pi(v) \leftarrow NIL$
	
	$T(v) \leftarrow \emptyset$
}

\For{each node $u \in N$}
{
	$d(u) \leftarrow 0$
	
	$T(u) \leftarrow T(u) \cup \{(u, 0)\}$
}

\KwRet $(d, \pi, T)$

\end{small}}
\end{algorithm}

\begin{algorithm}[!htb]
\caption*{\lrg{F}$AST$-\lrg{E}$ND$-\lrg{S}$TOPS\ (G, w, Q)$ \textbf{cont.}}
\label{algo212}
\func{\lrg{M}$ERGE\ $}{$u,v,c,d,\pi,T$}{
\begin{small}

$upd \leftarrow False$

$q_{prev}(v) \leftarrow |T(v)|$

\For{each query-cost pair $(p_{u},l_{u}) \in T(u)$}
{
	\If{$\exists$query-cost pair$(p_{v},l_{v}) \in T(v) \quad (p_{v}=p_{u})$}
	{
		\If{$l_{u}+c<l_{v}$}
		{
			$upd \leftarrow True$
		
			$l_{v} \leftarrow l_{u}+c$
		}
	}
	\Else
	{
		$upd \leftarrow True$	
	
		$T(v) \leftarrow T(v) \cup \{(p_{u},l_{u}+c)\}$
	}
}

\If{$upd = True$}
{
	$d_{temp}(v) \leftarrow \sum\limits_{(p_{v},l_{v}) \in T(v)} l_{v}$
	
	\If{$(|T(v)| > q_{prev}(v)) \lor (d_{temp}(v) < d(v))$}
	{
		$d(v) \leftarrow d_{temp}(v)$
		
		$\pi(v) \leftarrow NIL$
	}
}

\KwRet $(upd,d,\pi,T)$

\end{small}}
\end{algorithm}

Algorithm~\ref{algo2} outlines the pseudo-code for our novel solution approach. Line 1 computes the transpose graph $G^{T}$ of $G$, and the transpose cost function $w^{T}$ of $w$. Line 2 calls the \lrg{G}$ROUP$-\lrg{Q}$UERY$-\lrg{S}$EARCH$ routine that performs the search from the query source nodes. By a similar call to the same procedure, line 3 runs the search from the query destinations in the transpose graph $G^{T}$.

Lines 11-27 demonstrate the \lrg{G}$ROUP$-\lrg{Q}$UERY$-\lrg{S}$EARCH$ function. Line 12 initializes the search by calling \lrg{I}$NIT$-\lrg{G}$ROUP$-\lrg{Q}$UERIES$, shown in lines 35-43. For each node $v \in V$, the \textbf{for} loop in lines 36-39 sets its key cost $d(v)$ to $\infty$, its parent $\pi(v)$ to $NIL$, and the list of currently-reaching-queries $T(v)$ to $\emptyset$. The \textbf{for} loop in lines 40-42 initializes the cost $d(u)$ of each query node $u$ to $0$, and the list of currently-reaching-queries at $u$ to the corresponding query with cost $0$. After returning from the routine at line 43, line 13 builds the priory queue $PQ$ with all the nodes in the graph. Each time through the \textbf{while} loop of lines 14-26, line 15 extracts a vertex $u$ of the minimum key cost from $PQ$. Then, the \textbf{for} loop in lines 16-26 performs the relaxation to each node $v$ adjacent to $u$ and updates $d(v)$, $\pi(v)$, and $T(v)$ if required.

First, the call to the \lrg{M}$ERGE$ procedure in line 17 updates the shortest path costs from the individual query nodes. Lines 44-60 depict the \lrg{M}$ERGE$ routine. Line 45 initializes the variable $upd$, which denotes whether the function updates any query cost. Line 46 stores the number of queries currently reaching at $v$ before any update. Lines 47-54 update the query costs in $T(v)$ by considering a path through $u$ as an option; they also insert the newly-reaching-queries in $T(v)$. Second, we consider the $options\ 1~3$ as described above in computing the key cost of $v$, which determines its relative position in $PQ$. Lines 55-59 inside the \lrg{M}$ERGE$ function regard $option\ 1$, and $option\ 3$. After returning from the function in line 60, lines 18-21 deliberate $option\ 2$. Last, 22-26 update the search frontier represented by $PQ$. Line 27 returns the information acquired by the search.

The first search computes the key costs $d_{s}$, the parents $\pi_{s}$, and the individual query costs $T_{s}$. Similarly, The second search calculates the key costs $d_{d}$, the parents $\pi_{d}$, and the individual query costs $T_{d}$. After both searches terminate, lines 4-8 determine the optimal solution cost $opt$ by considering each node $v \in V$. They compute the candidate solution costs $d_{s}(v)+d_{d}(v)$, provided that both $T_{s}(v)$, and $T_{d}(v)$ contain all the queries, and find the best middle point $mid$ such that $opt = d_{s}(mid)+d_{d}(mid)$ is the optimal cost. Through a call to \lrg{C}$OMPUTE$-\lrg{E}$ND$-\lrg{S}$TOPS$ function, illustrated in lines 28-34, line 9 computes the optimal pair of end-stops $(st,en)$. Starting at $mid$, the function traverses the parent list $\pi_{s}$ (resp. $\pi_{d}$) until reaching $NIL$ to determine the start stop $st$ (resp. the end-stop $en$). Finally, line 10 returns the computed end-stops.

\subsubsection{Complexity Analysis}
\label{our_complexity}
In Algorithm~\ref{algo2}, line 1 requires $O(n+m)$ time to compute the transpose graph. Lines 2-3 make calls to the \lrg{G}$ROUP$-\lrg{Q}$UERY$-\lrg{S}$EARCH$ routine. We will discuss the time complexity of this procedure shortly. Lines 4-8 need $O(n)$ time to determine the optimal cost. The function call in line 9 requires another $O(n)$ time to compute the optimal end-stops.

Inside the \lrg{G}$ROUP$-\lrg{Q}$UERY$-\lrg{S}$EARCH$ function, lines 12-13 take $O(n+q)$ time to initialize the search and the priority queue. Determining the execution time of the \textbf{while} loop of lines 14-26 demand some rigorous analysis. We use $Fibonacci\ Heap$ to implement the min-priority queue $PQ$. Hence, the amortized time complexity of each \lrg{E}$XTRACT$-\lrg{M}$IN$ operation is $O(\lg n)$, and that of each \lrg{U}$PDATE$-$KEY$ or \lrg{I}$NSERT$ operation is $O(1)$. The call to the \lrg{M}$ERGE$ function in line 17 requires $O(q)$ time. Thus, each relaxation from a node $u$ to a node $v$, in lines 17-26, takes $O(q)$ time. The difficulty of this complexity analysis is in determining the number of \lrg{E}$XTRACT$-\lrg{M}$IN$ operations and the number of edge relaxations. Consider a relaxation from a node $u$ to a node $v$. For some queries waiting at $u$, we may find smaller costs of reaching $v$ through $u$, while for the other queries, the former costs of reaching $v$ may remain unchanged. Until all queries reach the node $v$, we compute its key cost as the summation of the costs of the currently-reaching-queries. Notice that after one of these unconventional relaxations, it is possible that the key cost of $v$ may become less than the key cost of $u$. However, for any particular query, we either improve or keep its cost, each time we relax that query along an edge. Nevertheless, the order of relaxation for any individual query may not be optimal, since we determine the relative order of a node in $PQ$ by its key cost, not by any query cost. Hence, our search procedure may not only insert a newly-reaching-query at a node $v \in V$ but also update the cost of a formerly-reaching-query, even after it extracts $v$ from $PQ$. Thus, we may insert and extract a node again and again. Be that as it may, after we relax a query at most $n-1$ times along each edge of the graph, further relaxations do not update its individual cost any more, like in the $Bellman$-$Ford$'s algorithm \cite{DBLP:books/daglib/0023376}. There are $q$ queries in total. Hence, we may need to extract a node at most $O(nq)$ times, giving $O(n^{2}q)$ number of \lrg{E}$XTRACT$-\lrg{M}$IN$ operations. Similarly, the maximum number of edge relaxations is $O(nmq)$. Therefore, the execution time of the search procedure is $O(n^{2}q \lg n + nmq^{2})$.

As the run time of the \lrg{G}$ROUP$-\lrg{Q}$UERY$-\lrg{S}$EARCH$ routine dominates the complexity of our algorithm, the worst-case time complexity of our approach is $O(n^{2}q \lg n + nmq^{2})$. Apparently, this bound is even worse that the $O(n^{2} \lg n)$ of the baseline algorithm. However, the expected performance of our technique outsmarts the baseline approach for the following reasons.

\begin{enumerate}
\item Since the sources (resp. the destinations) are co-located, the path-coherence property of the underlying road network ensures the existence of shared routes among the queries. Hence, we may usually relax multiple queries at once along an edge.
\item Although our search procedure does not guarantee the optimal relaxation order for any individual query, it provides a 'good order' for each query. Thus, in practice, the required number of relaxations is nearer to the best case of the $Bellman$-$Ford$'s algorithm than the worst case.
\end{enumerate}

Therefore, the number of relaxations along each edge is $\theta(q)$, rather than $O(nq)$, in practical circumstances. Thus, the expected time complexity of our algorithm is $O(nq \lg n + mq^{2})$, when executed on a road network with co-located sources, and destinations.

The adjacency list representations of edges in $G$ and $G^{T}$ need $O(n+m)$ space. For each node $v \in V$, our algorithm takes $O(1)$ space to store $d(v)$, $\pi(v)$, and a position in $PQ$; it requires another $O(q)$ space to store $T(v)$. Therefore, the overall space complexity of our approach is $O(m+nq)$.

\subsubsection{Improvement}
\label{imp_a2}
We have achieved further improvement upon Algorithm~\ref{algo2} by adopting the following pruning strategy. Instead of executing the two searches, in lines 2-3, separately, we run them in parallel. Unlike the original algorithm, we initialize $opt$ and $mid$ before the searches; we also move the \textbf{for} loop of lines 5-8 inside each search. During the searches, we continually obtain candidate values of $opt$. We terminate the first search after extracting a node $u$ from $PQ$, if $d_{s}(u)>=opt$. Similarly, we end the second search at an extracted node $u$, when $d_{d}(u)>=opt$.

\section{The Optimal Route and Intermediate Stops}
\label{oris}
In this section, we study the problem of finding the optimal route and the intermediate stops of a vehicle. First, we provide an optimal solution of exponential time and space complexities. Second, we propose our heuristic algorithm that achieves a near-optimal solution and requires polynomial time and space. Third, we analyze several variants of this problem and offer modifications of our original algorithm that solve the variants with similar efficiency and accuracy. Last, we show that the problem of finding the route-and-stops is a generalization of the traveling salesman path problem $(TSPP)$.

\subsection{Exact Solution}
\label{opt}
We compute the exact answer using the $Dijkstra$'s algorithm with bit-masking. We define each sub-problem (i.e., each state of the $Dijkstra$'s search) by a node and a subset of the query sources-and-destinations served by the vehicle along its path from the start-stop. In the accompanying subset of query nodes, which we represent using bit-masking in our implementation, a source stands for a passenger who has already entered the vehicle, while a destination corresponds to a user already dropped off on the path from the start-stop to the current node. During the progress of our algorithm, the search space comprises a collection of node-bitmask pairs, i.e., sub-problems, as defined above.

First, we initialize our algorithm by computing and storing the shortest path costs from the query sources (resp. all nodes) to all nodes (resp. the query destinations). By all nodes, we mean the nodes in the reduced graph (as discussed in Section~\ref{ov}). Second, we compute the optimal solution cost by performing the $Dijkstra$'s search technique. At each iteration of our search, we expand the search space by choosing a node-bitmask pair of the minimum cost and relaxing from that pair. We perform two types of relaxation. One is to branch to each neighbor node keeping the associated subset of sources-and-destinations fixed. The other is to grow the subset bitmask while remaining on the same node. We grow the subset of queries by adding either a new source, i.e., take a passenger onboard, or a new appropriate destination, i.e., drop one user off the vehicle. We use the edge costs and the costs between graph vertices and query nodes in the relaxation methods. Once a search-path reaches the end-stop with a full bitmask, it symbolizes that the vehicle has served all the users, and reached the final stoppage. By the end of the search, we have computed the optimal cost and the parent information for each state. Last, we calculate the optimal route and the intermediate stops en-route from the parent information computed during the search procedure.

\begin{figure}[!tb]
\centering
\includegraphics[width=.70\columnwidth]{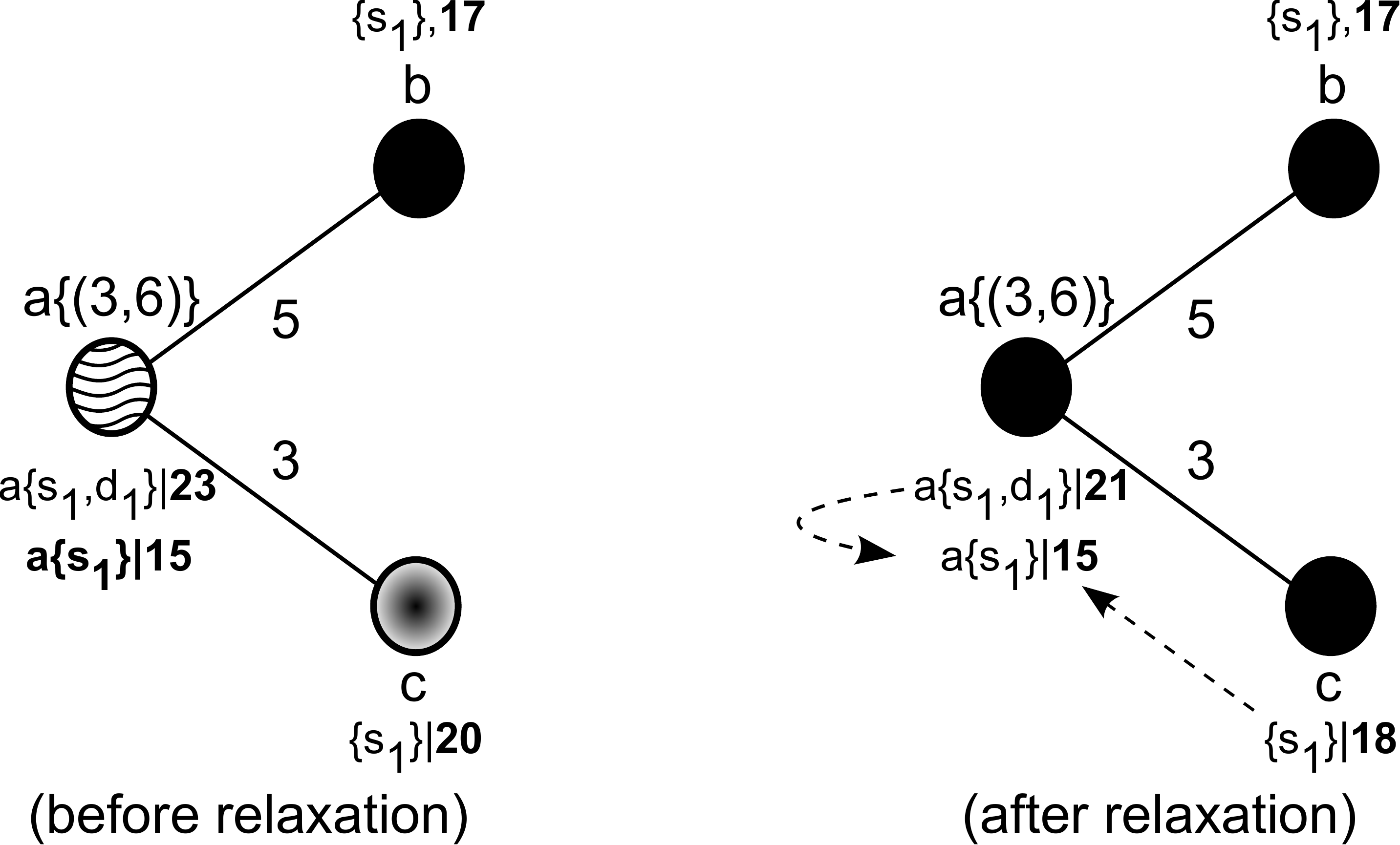}
\caption{An Example Relaxation Step of Algorithm~\ref{algo3}.}
\label{fig_algo3}
\end{figure}

Figure~\ref{fig_algo3} demonstrates an example relaxation step of the optimal algorithm; we show only the information necessary for our discussion. Node $a$ is $3$ units away from $s_{1}$, and $6$ units from $d_{1}$. Before relaxation, the cost at node $a$, with only $s_{1}$ served, is $15$; node $b$ (resp. $c$) has similar cost $17$ (resp. $20$). Also, the cost at node $a$, with both $s_{1}$ and $d_{1}$ served, is $23$ before relaxation. After relaxation from $a$, with $s_{1}$, $b$ remains unchanged $(15+5>17)$. We update the cost at $c$, with $s_{1}$, to $18$ $(15+3<20)$; $a$, with $s_{1}$, becomes the parent of $c$, with $s_{1}$. We also update the cost at $a$, with both $s_{1}$ and $d_{1}$, to $21$ $(15+6<23)$ and make $a$, with $s_{1}$ its parent. We update the search frontier $PQ$ accordingly.

\begin{algorithm}[!htb]
\caption{\lrg{O}$PT$-\lrg{S}$TOPS\ (G, w, Q, st, en)$}
\label{algo3}
\begin{small}
\setcounter{AlgoLine}{0}

\KwIn{A graph $G$, a cost function for edges $w$, a set of queries $Q$, the start-stop $st$, and the end-stop $en$}
\KwOut{An optimal sequence of stopping points $P=[P(1)=st,P(2),...,P(t-1),P(t)=en]$}

$Compute\ G^{T}$, and $w^{T}$

\For{each query $(s_{i},d_{i}) \in Q$}
{
	$SPC(s_{i},G.V) \leftarrow $ \lrg{D}$IJKSTRA\ (G, w, s_{i})$
	
	$SPC^{T}(d_{i},G.V) \leftarrow $ \lrg{D}$IJKSTRA\ (G^{T}, w^{T}, d_{i})$	
}

$(d,\pi) \leftarrow $ \lrg{I}$NIT$-\lrg{S}$INGLE$-\lrg{S}$OURCE\ (G.V \times \mathcal{P}(Q.S \cup Q.D), (st,\emptyset))$

$PQ \leftarrow G.V \times \mathcal{P}(Q.S \cup Q.D)$

\While{$PQ \neq \emptyset$}
{
	$(u,A) \leftarrow $ \lrg{E}$XTRACT$-\lrg{M}$IN\ (PQ)$
	
	\For{each source or destination node $r \in (Q.S \setminus A) \cup \{d_{i}: ((s_{i},d_{i}) \in Q)\land (s_{i} \in A)\land (d_{i} \notin A) \}$}
	{
		Let $D=SPC(r,u)$ (resp. $D=SPC^{T}(r,u)$), when $r \in Q.S$ (resp. $r \in Q.D$)
	
		$B \leftarrow A \cup \{r\}$
	
		$(d, \pi, PQ) \leftarrow$ \lrg{R}$ELAX\ ((u,A), (u,B), D, d, \pi, PQ)$
	}
	
	\For{each node $v \in G.Adj[u]$}
	{
		$(d, \pi, PQ) \leftarrow$ \lrg{R}$ELAX\ ((u,A),(v,A),w(u,v), d, \pi, PQ)$
	}
}

$P \leftarrow $ \lrg{C}$OMPUTE$-\lrg{S}$TOPS\ (st, en, \pi, Q)$

\KwRet $P$

\end{small}
\end{algorithm}

\begin{algorithm}[!htb]
\caption*{\lrg{O}$PT$-\lrg{S}$TOPS\ (G, w, Q, st, en)$ \textbf{cont.}}
\label{algo31}
\func{\lrg{I}$NIT$-\lrg{S}$INGLE$-\lrg{S}$OURCE\ $}{$S, u$}{
\begin{small}

\For{each state $s \in S$}
{
	$d(s) \leftarrow \infty$
	
	$\pi(s) \leftarrow NIL$
}

$d(u) \leftarrow 0$

\KwRet $(d,\pi)$

\end{small}}
\end{algorithm}

\begin{algorithm}[!htb]
\caption*{\lrg{O}$PT$-\lrg{S}$TOPS\ (G, w, Q, st, en)$ \textbf{cont.}}
\label{algo32}
\func{\lrg{R}$ELAX\ $}{$u, v, c, d, \pi, PQ$}{
\begin{small}

\If{$d(u)+c<d(v)$}
{
	$d(v) \leftarrow d(u)+c$
	
	$\pi(v) \leftarrow u$
	
	\lrg{D}$ECREASE$-$KEY\ (PQ, v, d(v))$
}

\KwRet $(d, \pi, PQ)$

\end{small}}
\end{algorithm}

\begin{algorithm}[!htb]
\caption*{\lrg{O}$PT$-\lrg{S}$TOPS\ (G, w, Q, st, en)$ \textbf{cont.}}
\label{algo33}
\func{\lrg{C}$OMPUTE$-\lrg{S}$TOPS\ $}{$st, en, \pi, Q$}{
\begin{small}

$P \leftarrow [en]$, $v \leftarrow en$, $B \leftarrow Q.S \cup Q.D$

\While{$v \neq st$}
{
	$(u,A) \leftarrow \pi(v,B)$
	
	\If{$(u=v) \land (v \notin \mathbf{elems}~P)$}
	{
		$P \leftarrow [v] + P$
	}
	
	$(v,B) \leftarrow (u,A)$
}
	
\If{$st \notin \mathbf{elems}~P$}
{
	$P \leftarrow [st] + P$
}

\KwRet $P$

\end{small}}
\end{algorithm}

Algorithm~\ref{algo3} demonstrates the pseudo-code for the optimal solution approach. Lines 1-4 compute and store the shortest path costs between the query nodes and the graph vertices. Line 5 calls \lrg{I}$NIT$-\lrg{S}$INGLE$-\lrg{S}$OURCE$ to initialize the search. Lines 17-22 show the function, which sets the parents to $NIL$ and the costs to $\infty$, except that of the start state, which it initializes to $0$. Each state is a node-bitmask pair from the cartesian product $G.V \times \mathcal{P}(Q.S \cup Q.D)$. Line 6 initializes the min-priority queue $PQ$ to contain all the states in the product. In each iteration of the \textbf{while} loop of lines 7-14, line 8 extracts the state $(u,A)$ of minimum cost from among the states in $PQ$. Then, lines 9-12 choose each suitable query node $r$ such that $r \notin A$ and compute the cost $D$ between $r$ and $u$, and the set $B = A \cup \{r\}$; then, relax to $(u,B)$ from $(u,A)$ with cost $D$. Subsequently, the \textbf{for} loop of lines 13-14 considers each adjacent node $v$ of $u$ and relaxes to $(v,A)$ from $(u,A)$ with cost $w(u,v)$. Lines 23-28 depict the \lrg{R}$ELAX$ procedure, which minimizes the cost of the destination state and updates the parent, and the priority queue if necessary. Finally, line 15 calls the \lrg{C}$OMPUTE$-\lrg{S}$TOPS$ function, which we show in lines 29-38; it computes the optimal sequence of intermediate stops $P$ using the parent information $\pi$. Line 16 returns $P$, and the algorithm terminates.

\subsubsection{Complexity Analysis}
\label{opt_complexity}
In Algorithm~\ref{algo3}, time complexity of lines 1-4 is $O(q(n \lg n+m))$. Line 5 takes $O(total\ number\ of\ states)$, which amounts to $O(n|\mathcal{P}(Q.S \cup Q.D)|)$. At first glance, the size of the power set seems to be $O(2^{2q})$. However, not all subsets of query sources-and-destinations are valid. Clearly, the subsets that contain a query destination without its corresponding source are not reachable from the start position. A nice implementation would consider only the valid bitmasks and reduce the total number of subsets to $O(3^{q})$. Therefore, line 5 requires $O(n3^{q})$. Inside the \textbf{while} loop of lines 7-14, we extract a node-bitmask pair exactly once. Thus, the total number of \lrg{E}$XTRACT$-\lrg{M}$IN$ operations performed in line 8 is $O(n3^{q})$. Similarly line 12 calls the \lrg{R}$ELAX$ routine $O(n3^{q})$ times. However, line 14 executes $O(m3^{q})$ times throughout the \textbf{while} loop, since for each subset, our algorithm relaxes along an edge exactly once. Hence, the total number of \lrg{D}$ECREASE$-$KEY$ operations performed is $O((n+m)3^{q})$. When we use $Fibonacci\ Heap$, the amortized cost of each \lrg{E}$XTRACT$-\lrg{M}$IN$ operation is $O(\lg (n3^{q}))$, i.e., $O(q \lg n)$ and each \lrg{D}$ECREASE$-$KEY$ operation is $O(1)$. Therefore, the time complexity of the \textbf{while} loop is $O((n+m)3^{q}+nq3^{q} \lg n)$. Finally, line 15 executes in $O(n3^{q})$ time. The complexity of the \textbf{while} loop dominates in the analysis. Usually, $nq \lg n >> m$. Consequently, the overall time complexity of Algorithm~\ref{algo3} is $O(nq3^{q} \lg n)$.

The Adjacency List representation of the road network requires $O(n+m)$ space. The search space takes $O(n3^{q})$. Therefore, the overall space complexity of the optimal algorithm is $O(m+n3^{q})$.

\subsubsection{Improvement}
\label{imp_a3}
Instead of waiting for $PQ$ to be empty, we may terminate the search early, after extracting the pair $(en,Q.S \cup Q.D)$ from $PQ$. We may achieve further improvement by pruning the search space with a heuristic estimate such as the one in Section~\ref{app2}.

\subsection{Heuristic Solution}
\label{app2}
The exact solution approach, presented in Section~\ref{opt}, has both time and space complexities exponential in the number of queries. Therefore, it works for only a small group of users. What we need is a more scalable algorithm that can expeditiously produce a solution for a large number of passengers. In this section, we present a novel heuristic algorithm that efficiently computes a near-optimal answer. In this approach, unlike the optimal method, we do not explore an exponential number of sub-problems. Rather, we make greedy selections to keep the size of our search space within a polynomial bound; this results in an algorithm with polynomial time and space complexities. However, we lose a little accuracy in the process. Our approach provides a near-optimal solution with a reasonably small error and works well in practice.

In our heuristic, for each node $v \in V$ in the reduced road network graph (as discussed in Section~\ref{ov}), we aim at finding a route of the vehicle from the start-stop $st$, which minimizes the cost function $C_{2}$. For many a node, we only manage to reach a sub-optimal solution by our greedy technique. Throughout our search, we keep a frontier of nodes, from where we are yet to branch to their adjacent nodes. We maintain the search frontier using a priority queue $PQ$; we order the nodes in $PQ$ by their costs. In the beginning, $PQ$ contains only the start-stop $st$, with each passenger both entering and exiting the vehicle at $st$. Then, we gradually expand our search space by each time extracting a node with the minimum cost from $PQ$ and greedily relaxing to its adjacent nodes. By relaxation, we mean the process of branching the search to a neighbor node. The route of the vehicle to a node $u$ is a sequence of stops, $P=[P(1)=st,P(2),...,P(t-1),P(t)=u]$; each passenger gets on at some stop $P(i)$, and off at another stop $P(j)$, with $i<=j$. The cost of such a node $u$ is $C_{2}(P)$, i.e., a summation of the vehicle's route cost to $u$ and the passengers' solo travel costs to or from the vehicle. Suppose, at one point of our algorithm, the node $u$ is the frontier node of the least cost. We extend the search by relaxing from $u$ to each of its neighbor nodes $v \in G.Adj[u]$. During the relaxation to a node $v$ from a node $u$, we consider the following options for each passenger:

\begin{enumerate}
\item S/he may get on and off the vehicle on the path from the start-stop to $u$.
\item S/he may enter at or before $u$ and exit at $v$.
\item S/he may both enter and exit at $v$.
\end{enumerate}

We determine the cost $C_{2}(P)$, $P=[P(1)=st,P(2),...,P(t-1)=u,P(t)=v]$, by taking the minimum of the above three choices for each passenger. If $C_{2}(P)$ is less than the current best at $v$, we assign $v$'s new cost to $C_{2}(P)$ and its parent to $u$; we also update the passengers' lone travel costs outside the vehicle accordingly. We will illustrate the details with an example shortly.

Remember, we grow the search space by each time picking the frontier node of the minimum cost and relaxing to its neighbors. Notice that two factors are affecting the cost of a relaxation to a node $v$ from a node $u$. One is the positive edge cost $w(u,v)$ between $u$ and $v$. The other is a possible decrease in cost as a result of the availability of better location options for stops, with $v$ as an additional choice, as described above. Due to a balance between these two opposing factors, the cost of a relaxation may be either positive or negative. Suppose, at one time, we have relaxed from a node $u$. Later, due to the existence of negatively weighted relaxations, we may find a better route to reach $u$. Depending on whether or not we allow relaxing to such a node $u$, we may implement two versions of our algorithm. In one version, we allow relaxing to a node, already extracted from the search frontier. In this version, we may even encounter negative weighted cycles. However, we do not require traversing a cycle more than twice; we do not gain any new information when traversing for the third time, and the cost of the vehicle's route keeps increasing after each relaxation. In the other version, like the $Dijkstra$'s algorithm, we do not permit relaxing to an already extracted node. The former version is slightly more accurate and less efficient than the latter. Shortly, we will illustrate with an example that both versions fail to compute the optimal solution. For simplicity, we will present only the latter version in Algorithm~\ref{algo4}. We execute the search exhaustively until the search frontier $PQ$ becomes empty. Then, we compute our answer sequence of stops between the start-stop $st$ and the end-stop $en$ from the stored parent information and report as the outcome of our algorithm.

\begin{figure}[!tb]
\centering
\includegraphics[width=.90\columnwidth]{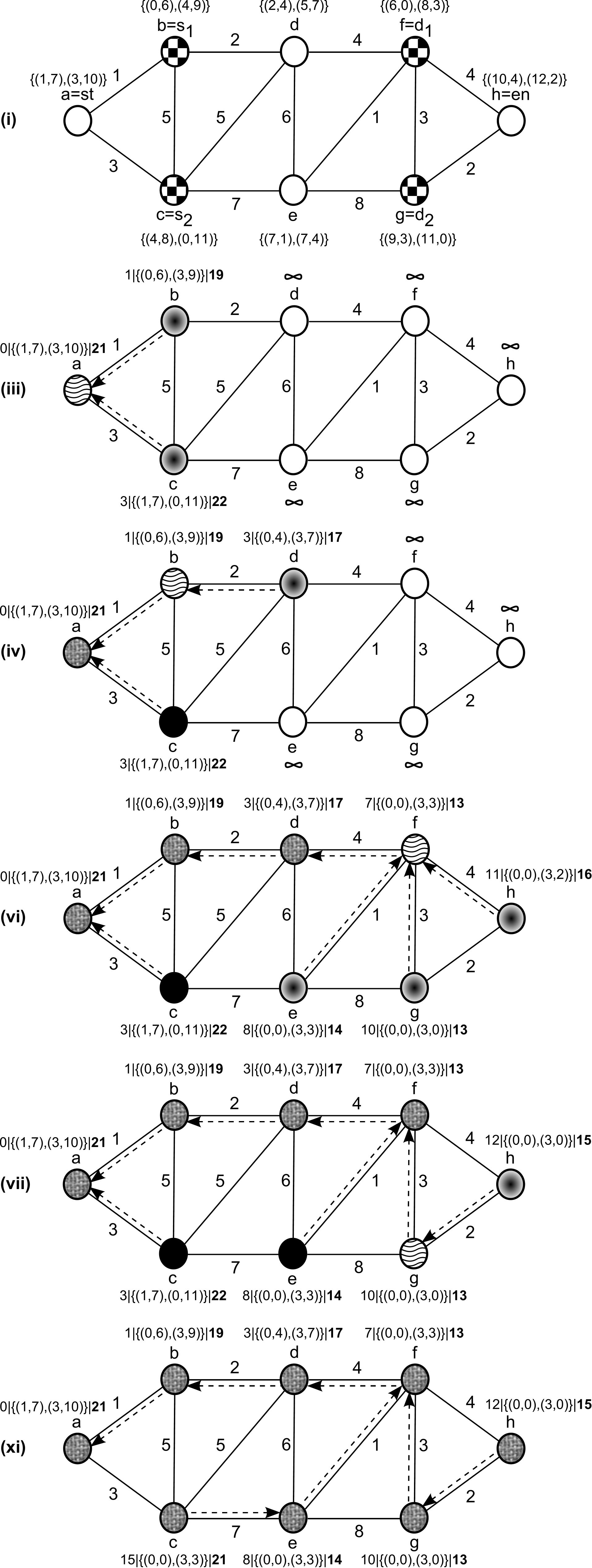}
\caption{Example Phases of Algorithm~\ref{algo4} - \\Steps (i), (iii), (iv), (vi), (vii), and (xi).}
\label{fig_algo4}
\end{figure}

We illustrate our heuristic search procedure with an example in Figure~\ref{fig_algo4}. In this figure, we show several steps of our algorithm in a sample graph with artificial queries. Query $1$ is from node $b$ to node $f$, while query $2$ is from node $c$ to node $g$. Node $a$ (resp. node $h$) is the vehicle's start-stop (resp. end-stop). In the initialization phase, step $(i)$, we compute the shortest path cost between each node and each query node. For example, the four numbers in the entry $\{(2,4),(5,7)\}$ at node $d$ indicate the smallest travel costs between $d$ and the query nodes - $s_{1}$, $d_{1}$, $s_{2}$, and $d_{2}$ respectively. We omit step $(ii)$ from the figure for brevity, where node $a$ is the sole member of the search frontier $PQ$. In step $(iii)$, we relax from $a$ to $b$, and $c$. Let us clarify the entry beside each node; consider $0|\{(1,7),(3,10)\}|21$ at node $a$. The first number, $0$, designates the cost of the vehicle's route from $st$ to $a$. The second number, $1$, shows the cost of travel of the passenger in $Query\ 1$ before entering the vehicle, while the third number, $7$, demonstrates the path cost of the same passenger after exiting the vehicle. Similarly, the fourth and the fifth numbers, $3$ and $10$, depict the solo travel costs of the passenger in $Query\ 2$ outside the vehicle. The last number, $21$, indicates the total cost, which is a summation of the first five numbers. The vehicle moves to $b$ (resp. $c$) with cost $1$ (resp. $3$). At $b$, for query $1$, the cost-pair $(0,6)$ dominates over the pair $(1,7)$, i.e., the first passenger both enters and exits at $b$; for query $2$, the second passenger gets on at $a$ and off at $b$, ensuring the cost-pair $(3,9)$ to prevail. At $c$, $(1,7)$ triumphs over $(4,8)$ for query $1$, i.e., passenger $1$ both boards and leaves at $a$; for query $2$, $(0,11)$ persists. Note that $(3,10)$, $(0,11)$, and $(3,11)$ are the valid choices for the second passenger, among which $(0,11)$ is the minimum; $(0,10)$ is invalid, since a passenger cannot leave a vehicle before entering. Node $a$ becomes the parent of each $b$, and $c$. We push $b$ and $c$ to $PQ$ and mark $a$'s costs as final. Then, in step $(iv)$, we relax from $b$ to its neighbors, except $a$. Node $c$'s costs and status remain unchanged. Node $d$ enters the search frontier $PQ$, $b$ being its parent. In step $(v)$ that we leave out, we relax from $d$ to its adjacent nodes. After that in step $(vi)$, we relax from $f$; it becomes the parent of $e$, $g$, and $h$. Notice again, once we have relaxed from a node, such as $d$, we never relax to it throughout the remainder of our search. In step $(vii)$, after relaxing from $g$, it becomes $h$'s parent, replacing $f$. We exclude the next three steps. Finally, step $(xi)$ depicts the outcome of our algorithm. We determine the vehicle's route as $a-b-d-f-g-h$ from the parent information. Each passenger gets on the vehicle at a node in its route, nearest from her/his source; similarly, s/he gets off at a node nearest to her/his destination. We report a node in the vehicle's route as a stoppage if at least one user enters or exits at that node. In our example, user $1$ gets on at $b$ and off at $f$; user $2$ boards at the start-stop $a$ and leaves at $g$. Therefore, $b$, $f$, and $g$ are our vehicle's intermediate stops.

\begin{figure}[!tb]
\centering
\includegraphics[width=.80\columnwidth]{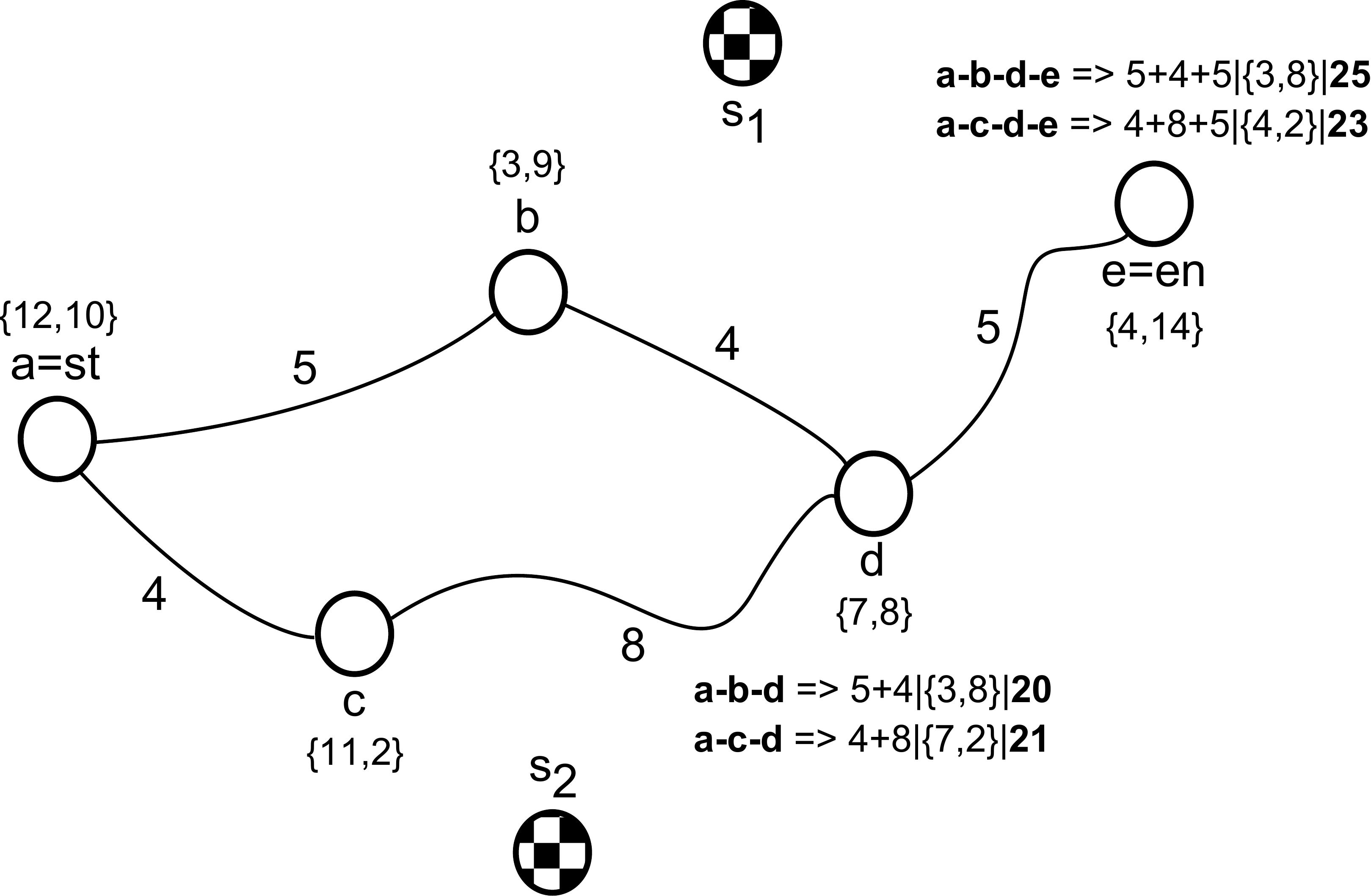}
\caption{An Example Demonstrating the Sub-optimality of Algorithm~\ref{algo4}.}
\label{fig2_algo4}
\end{figure}

In the above instance, our heuristic search produces the optimal solution. The next example in Figure~\ref{fig2_algo4} illustrates that our algorithm does not always produce the optimal answer. For brevity, we leave out any destination and consider only two source nodes, $s_{1}$ and $s_{2}$, in this example. In the figure, we have also omitted any node, edge, or path irrelevant to our discussion. $a$ is the start-stop, and $e$ is the end-stop. The entry at each node shows its distances from the two sources, e.g., $\{11,2\}$ at node $c$ means that $c$ is $11$ units distant from $s_{1}$, and $2$ units from $s_{2}$. First, consider the paths $a-b-d$ and $a-c-d$. The former costs $20\ (5+4|\{3,8\}|20)$, while the latter costs $21\ (4+8|\{7,2\}|21)$. Hence, our algorithm keeps the former path and forgets the latter. Then, it greedily proceeds to calculate the path $a-b-d-e$ ending at $e$, whose cost is $25\ (5+4+5|\{3,8\}|25)$. Our search technique never considers the path $a-c-d-e$ ending at $e$, whose cost is $23\ (4+8+5|\{4,2\}|23)$. Thus, the greedy choice made at node $d$ has prevented our heuristic from computing the optimal answer, and we have only managed to reach a sub-optimal solution. Be that as it may, while at node $d$, we had no other choice than to forget the more expensive path. Of course, we would succeed in computing the optimal answer by remembering every path; however, such an algorithm would have exponential time complexity. We have chosen to provide a polynomial-time algorithm at the price of a little accuracy.

\begin{algorithm}[!htb]
\caption{\lrg{H}$EUR$-\lrg{S}$TOPS\ (G, w, Q, st, en)$}
\label{algo4}
\begin{small}
\setcounter{AlgoLine}{0}

\KwIn{A graph $G$, a cost function for edges $w$, a set of queries $Q$, the start-stop $st$, and the end-stop $en$}
\KwOut{An optimal sequence of stopping points $P=[P(1)=st,P(2),...,P(t-1),P(t)=en]$}

$Compute\ G^{T}$, and $w^{T}$

\For{each query $(s_{i},d_{i}) \in Q$}
{
	$SPC(s_{i},G.V) \leftarrow $ \lrg{D}$IJKSTRA\ (G, w, s_{i})$
	
	$SPC^{T}(d_{i},G.V) \leftarrow $ \lrg{D}$IJKSTRA\ (G^{T}, w^{T}, d_{i})$	
}

$(d, \pi, T) \leftarrow $ \lrg{I}$NIT$-\lrg{V}$EHICLE\ (G, Q, st, SPC, SPC^{T})$

$PQ \leftarrow G.V$

\While{$PQ \neq \emptyset$}
{
	$u \leftarrow $ \lrg{E}$XTRACT$-\lrg{M}$IN\ (PQ)$
	
	$mark(u) \leftarrow True$
	
	\For{each node $v \in G.Adj[u]$}
	{
		\If{$mark(v) = False$}
		{
			$(d, \pi, T, PQ) \leftarrow$ \lrg{R}$ELAX$-\lrg{V}$EHICLE\newline(u, v, w(u,v), st, Q, SPC, SPC^{T}, d, \pi, T, PQ)$
		}
	}
}

$P \leftarrow $ \lrg{V}$EHICLE$-\lrg{S}$TOPS\ (st, en, \pi, T)$

\KwRet $P$

\end{small}
\end{algorithm}

\begin{algorithm}[!htb]
\caption*{\lrg{H}$EUR$-\lrg{S}$TOPS\ (G, w, Q, st, en)$ \textbf{cont.}}
\label{algo41}
\func{\lrg{I}$NIT$-\lrg{V}$EHICLE\ $}{$G, Q, st, C, C^{T}$}{
\begin{small}

\For{each node $v \in G.V$}
{
	$d(v) \leftarrow \infty$

	$\pi(v) \leftarrow NIL$
	
	$T(v) \leftarrow \emptyset$
	
	$mark(v) \leftarrow False$
	
	$stop(v) \leftarrow False$
}

$d(st) \leftarrow \sum\limits_{i=1}^{q} (C(s_{i}, st)+C^{T}(d_{i},st))$

$T(st) \leftarrow T(st) \cup \{(st,0)\}$

\For{each query $(s_{i},d_{i}) \in Q$}
{
	$T(st) \leftarrow T(st) \cup \{(s_{i}, C(s_{i},st)),(d_{i},C^{T}(d_{i},st))\}$
}

\KwRet $(d, \pi, T)$

\end{small}}
\end{algorithm}

\begin{algorithm}[!htb]
\caption*{\lrg{H}$EUR$-\lrg{S}$TOPS\ (G, w, Q, st, en)$ \textbf{cont.}}
\label{algo42}
\func{\lrg{R}$ELAX$-\lrg{V}$EHICLE\ $}{$u, v, c, st, Q, C, C^{T}, d, \pi, T, PQ$}{
\begin{small}

Let the vehicle-cost pair $(st,c_{v}) \in T(u)$
		
$d_{temp}(v) \leftarrow c_{v}+c$

$T_{temp}(v) \leftarrow \{(st, c_{v}+c)\}$

\For{each query $(s_{i},d_{i}) \in Q$}
{
	Let the query-cost pairs $(s_{i},c_{s}),(d_{i},c_{d}) \in T(u)$

	\If{$c_{s}+\min(c_{d},C^{T}(d_{i},v))<C(s_{i},v)+C^{T}(d_{i},v)$}
	{
		$d_{temp}(v) \leftarrow d_{temp}(v)+c_{s}+\min(c_{d},C^{T}(d_{i},v))$

		$T_{temp}(v) \leftarrow T_{temp}(v)\cup \{(s_{i},c_{s}), (d_{i},\min(c_{d},C^{T}(d_{i},v)))\}$
	}
	\Else
	{
		$d_{temp}(v) \leftarrow d_{temp}(v)+C(s_{i},v)+C^{T}(d_{i},v)$

		$T_{temp}(v) \leftarrow T_{temp}(v)\cup \{(s_{i},C(s_{i},v)), (d_{i},C^{T}(d_{i},v))\}$
	}
}

\If{$d_{temp}(v)<d(v)$}
{
	$d(v) \leftarrow d_{temp}(v)$

	$\pi(v) \leftarrow u$

	$T(v) \leftarrow T_{temp}(v)$
	
	\If{\large{E}\small$XISTS\ (PQ, v)$}
	{
		\lrg{D}$ECREASE$-$KEY\ (PQ, v, d(v))$
	}
	\Else
	{
		\lrg{I}$NSERT\ (PQ, v)$
	}
}

\KwRet $(d, \pi, T, PQ)$

\end{small}}
\end{algorithm}

\begin{algorithm}[!htb]
\caption*{\lrg{H}$EUR$-\lrg{S}$TOPS\ (G, w, Q, st, en)$ \textbf{cont.}}
\label{algo43}
\func{\lrg{V}$EHICLE$-\lrg{S}$TOPS\ $}{$st, en, \pi, T$}{
\begin{small}

Let $VP$ be the set of nodes in the vehicle's route from $st$ to $en$

\For{each query $(s_{i},d_{i}) \in Q$}
{
	Let $a$ (resp. $b$) be the nearest node to $s_{i}$ (resp. $d_{i}$) in $VP$; in case of ties, it is the closest node to $st$
	
	$stop(a) \leftarrow True,\ stop(b) \leftarrow True$
}

$P \leftarrow [en]$, $v \leftarrow \pi(en)$

\While{$v \neq st$}
{
	\If{$stop(v) = True$}
	{
		$P \leftarrow [v]+P$
	}

	$v \leftarrow \pi(v)$
}

$P \leftarrow [st]+P$

\KwRet $P$

\end{small}}
\end{algorithm}

Algorithm~\ref{algo4} provides the pseudo-code for our heuristic search algorithm. Lines 1-4 pre-compute the shortest path costs between the query nodes and the graph vertices. Line 5 calls \lrg{I}$NIT$-\lrg{V}$EHICLE$; the function, shown in lines 15-26, initializes the search. The \textbf{for} loop of lines 16-21 sets the cost $d(v)$ of each node $v \in V$ to $\infty$, the parent $\pi(v)$ to $NIL$, the initial list of passengers $T(v)$, served en-route from $st$ to $v$, to $\emptyset$, the relaxation status $mark(v)$ to $False$, and the stoppage track $stop(v)$ to $False$. Lines 22-25 initialize $d(st)$ and $T(st)$, assuming all the passengers both entering and exiting at $st$. After returning from the function at line 26, line 6 initializes the min-priority queue $PQ$ to contain all the vertices in $V$. Each time through the \textbf{while} loop of lines 7-12, lines 8-9 extract a vertex $u$ of the minimum cost from $PQ$ and mark it as extracted. Then, lines 10-12 relax each edge $(u,v)$ leaving $u$, except when $v$ is already extracted; thus, updating $d(v)$, $\pi(v)$, and $T(v)$, if we can improve the best sequence of stops ending at $v$ by going through $u$.

Lines 27-47 show the \lrg{R}$ELAX$-\lrg{V}$EHICLE$ routine. Lines 28-38 calculate the candidate cost update $d_{temp}(v)$ and passenger info update $T_{temp}(v)$. Lines 28-30 consider the vehicle going to $v$ from $u$, update the route cost by the edge cost, and insert an entry for the vehicle in $T_{temp}(v)$. For each user, lines 31-38 regard $v$ as a potential stop, check whether exiting or both entering and exiting at $v$ is less costly than the former choices, and update $d_{temp}(v)$ and $T_{temp}(v)$ accordingly. If $d_{temp}(v)$ is better than the current best estimate $d(v)$ of $v$, lines 39-46 set $d(v)$ to $d_{temp}(v)$, $\pi(v)$ to $u$, and $T(v)$ to $T_{temp}(v)$ and update the priority queue $PQ$ accordingly. We return from the procedure at line 47. After the search exhausts, line 13 computes the answer sequence of stops $P$ by calling \lrg{V}$EHICLE$-\lrg{S}$TOPS$. The function \lrg{V}$EHICLE$-\lrg{S}$TOPS$, shown in lines 48-59, marks the nearest node to each query node in the vehicle's route as a stoppage. Then, it traverses the parent list $\pi$ backwards starting from $en$ and builds the answer sequence of stops $P$. Finally, line 14 returns $P$, as computed by \lrg{V}$EHICLE$-\lrg{S}$TOPS$, as the answer of our algorithm.

\subsubsection{Complexity Analysis}
\label{heur_complexity}
In Algorithm~\ref{algo4}, lines 1-4 require $O(q(n \lg n+m))$ time. The procedure \lrg{I}$NIT$-\lrg{V}$EHICLE$ called in line 5 takes $O(n+q)$ time. Line 6 needs $O(n)$ time to initialize the min-priority queue $PQ$, implemented with a $Fibonacci\ Heap$. The \lrg{V}$EHICLE$-\lrg{S}$TOPS$ routine call in line 13 requires $O(nq)$ time to compute the answer sequence of stops from parent information.

Let us analyze the time complexity of the \textbf{while} loop of lines 7-12. In the version that we have presented in Algorithm~\ref{algo4}, we do not countenance relaxation to an already extracted vertex; lines 9 and 11 guarantee that. Hence, line 8 extracts each vertex $u \in V$ from $PQ$ exactly once. Similarly, we relax along each edge exactly once; a call to the \lrg{R}$ELAX$-\lrg{V}$EHICLE$ routine incurs $O(q)$ time. Again, we know that a $Fibonacci\ Heap$ implementation of $PQ$ requires $O(\lg n)$ time per \lrg{E}$XTRACT$-\lrg{M}$IN$ operation and $O(1)$ time per \lrg{I}$NSERT$ or \lrg{D}$ECREASE$-$KEY$ operation. Thus, the time complexity of the \textbf{while} loop is $O(n \lg n + mq)$. The overall time complexity of this version of our algorithm is $O(q(n \lg n+m))$.

In the other version, we permit relaxation to a previously extracted vertex, i.e., lines 9 and 11 are absent. In it, the complexity analysis of the \textbf{while} loop is a bit involved. Due to the existence of negatively weighted relaxations, unlike the $Dijkstra$'s algorithm, it is no longer guaranteed that line 8 extracts each vertex $u \in V$ from $PQ$ exactly once. We may, in fact, insert and extract a vertex again and again. We may even encounter negatively weighted cycles; however, such a cycle does not remain negative after traversing it twice. Hence, like the $Bellman$-$Ford$-$Moore$'s algorithm, we may require relaxing an edge at most $O(n)$ times. Similarly, we may extract a node at most $O(n)$ times. Therefore, the worst-case time complexity of this version of our algorithm is $O(n^{2} \lg n + nmq)$. However, when we use a priority queue to order the nodes in the search frontier, it reduces the total number of relaxations by greedily choosing a vertex of the least cost first. Besides, despite the presence of negatively weighted relaxations, the path cost of the vehicle increases in each relaxation. Thus, in practice, we relax each edge only a constant times on average. Therefore, the expected time complexity of this version reduces to $O(n \lg n + mq)$. In a road network, which is usually sparse, $m = O(n)$. Again, usually, $q = \omega (\lg n)$. Hence, the average computation time is only $O(nq)$.

The adjacency list representations of edges in $G$ and $G^{T}$ require $O(n+m)$ space. At each node $v \in V$, our algorithm requires $O(1)$ space to store $d(v)$, $\pi(v)$, $mark(v)$, and $stop(v)$, $O(q)$ to store $T(v)$, and another $O(q)$ for the shortest path costs $SPC(Q.S,v)$ and $SPC^{T}(Q.D,v)$. As a result, the overall space complexity of our heuristic solution approach is $O(m+nq)$.

\subsubsection{Improvement}
\label{imp_a4}
We have adopted the following pruning technique to improve upon Algorithm~\ref{algo4}. After we extract a node $u$ from $PQ$ in line 8, we check whether the vehicle's route cost from $st$ to $u$ exceeds $d(en)$; if positive, we simply terminate the search rather than waiting for $PQ$ to be empty.

\subsection{Variants}
\label{var}
In this section, we analyze the variants introduced in Section~\ref{pors}. We propose modifications of our algorithms to solve each variant.

\subsubsection{Constraint on Each User's Lone Path Length Before Entering or After Exiting}
\label{c1}
In this variant, we limit the maximum length a user may travel before entering, or after exiting the vehicle to $R_{1}$. To solve this variant, in both Algorithm~\ref{algo3} and Algorithm~\ref{algo4}, we replace each $SPC(s_{i},v)>R_{1}$, and $SPC^{T}(d_{i},v)>R_{1}$ with a very large number $LN \rightarrow \infty$. The other large number that we use in our implementations to represent $\infty$ should be at least $(2q+1)LN$. We keep each $SPC(s_{i},v)<=R_{1}$, and $SPC^{T}(d_{i},v)<=R_{1}$ as is. This way, we prevent our algorithms to compute a path for the vehicle, which does not meet the constraint, by increasing its cost. Our searches find the best possible routes conforming to the limitation; the complexities remain the same.

\subsubsection{Constraint on the Vehicle's Route Length}
\label{c2}
In our second variant, we restrict the vehicle's path cost to $R_{2}*SPC(st,en)$. In both algorithms, when relaxing from a node $u$ to an adjacent node $v$, if the vehicle's route becomes more costly than $R_{2}*SPC(st,en)$, we do not update the cost of $v$. Thus, each algorithm eventually calculates an answer, where the vehicle's path cost remains within the limit. The complexities remain the same as of the original algorithms.

\subsubsection{Constraint on the Entering/Exiting Passenger-Cardinality at a Stop}
\label{c3}
Here, we require that at least $R_{3}$ passengers get on-or-off the vehicle at each intermediate stop. In Algorithm~\ref{algo3}, notice that from a state $(u,A)$, we perform two types of relaxations. One to a state $(u,B)$, i.e., we grow the set of queries waiting at $u$ by an additional source or destination node; another to a state $(v,A)$, where $v$ is a neighbor of $u$. In the former type, we adopt the following modification. If the parent of $(u,A)$ is $(u',A)$, where $u \neq en$, we grow the set $A$ by $R_{3}$ new sources or destinations to form $B$ and relax to $(u,B)$; otherwise, we proceed as in our original algorithm. Doing this ensures that each intermediate stop serves at least $R_{3}$ passengers. The time and space complexities remain exponential in the number of queries.

In Algorithm~\ref{algo4}, recall that we keep the parent of each node, which facilitates the computation of our final sequence of stops, later on. To solve this variant, storing the parent information is not sufficient. At each node $v$, we maintain the vehicle's route from $st$ to $v$ as a sequence of stops $P$; we also keep a list of passengers entering/exiting at each stop. If upon a prospective relaxation to a node $v$, the passenger-cardinality of an intermediate stop is to drop below $R_{3}$, we greedily eliminate that stoppage and distribute each of its passengers to another stop, where his/her lone travel cost is the lowest. If this new relaxation to $v$ succeeds, we update its cost and other information accordingly. Our approach provides a near-optimal solution; the worst-case time and space complexities remain the same as of Algorithm~\ref{algo4}.

\subsubsection{Constraint on the Total Number of Stops}
\label{c4}
In the fourth variant, we directly restrict the total number of stops by demanding that it is less than or equal to $R_{4}$. There is no easy modification of Algorithm~\ref{algo3} that solves this version. We require an additional parameter in the state representation, namely, the number of stops. Therefore, we represent each state by a triplet $(u,A,t)$; $u$ is a node, $A$ is the set of passengers served by the vehicle in its path from $st$ to $u$, and $t$ is the number of stops in its route including $st$ and $u$. From $(u,A,t)$, we first relax to $(u,B,t)$. Then, for each neighbor $v$ of $u$, we relax to $(v,A,t)$ if $(u,A,t)$'s parent is $(u',A,t)$ or $(u',A,t-1)$; we relax to $(v,A,t+1)$ when $(u,A',t)$ is the parent of $(u,A,t)$. After the search terminates, the minimum among $d(en,U,2), d(en,U,3), ..., d(en,U,R_{4})$ is the cost of our solution. Finally, we build the result sequence of stops from the parent information. The overall time complexity of this approach is $O(nq^{2}3^{q} \lg (nq))$; the space complexity is $O(m+nq3^{q})$.

Like in the modification of Algorithm~\ref{algo4} for the third variant, at each node $v$, we keep a sequence of stops $P$ representing the vehicle's route and a list of passengers served at each stop. We consider each potential relaxation to see whether it causes the number of stops to become larger than $R_{4}$. In that circumstance, we greedily get rid of a stoppage of the lowest passenger-cardinality (or the lowest total passenger-costs, in the case of ties) and distribute each concerned passenger to another stop, where his/her individual travel cost is the least. If the new relaxation becomes successful, we update the costs and other information accordingly. Our approach finds a near-optimal sequence of stops conforming to the constraint and has the same worst-case complexities as of the original algorithm.

\subsubsection{The Weighted Version}
\label{c5}
In our last variant, we discriminate between the route cost of the vehicle and the total lone travel cost of the passengers by assigning them unequal weights, specifically, $R_{5}$ and $(1-R_{5})$ respectively. To solve this variant, in both algorithms, we compute the objective cost as defined in Section~\ref{pc5}; the time and the space complexities remain the same.

\bigskip
The nature of our solution approach to each variant permits the possibility of merging more than one of the above techniques to solve additional variants with multiple constraints.

\subsection{Relation with TSPP}
\label{tspp}
Notice that in the first variant of our second problem, introduced in Section~\ref{pc1}, we require that each passenger's individual path cost before entering or after exiting the vehicle is less than or equal to $R_{1}$. In our original problem, $R_{1} \rightarrow \infty$. Contrarily, when we set $R_{1}$ to $0$, we force our algorithms to compute a route of the vehicle that passes through each query node. Therefore, with $R_{1}=0$, our problem becomes the standard $TSPP$. Although this analysis does not rigorously prove the NP-hardness of the $ORIS$ problem, we may safely say that the $TSPP$ is a special case of the first variant of our $ORIS$ query. In other words, ours is more general than the $TSPP$. If we succeed in solving the first variant optimally in polynomial time for any $R_{1}$, we will have solved both the $ORIS$ and the $TSPP$.

Similarly, as \cite{DBLP:journals/tkde/LiQYM16} proves, the weighted version of our problem (Section~\ref{pc5}) reduces to the $TSPP$, when $R_{5}=0.\overline{3}$.

\section{Experimental Results}
\label{exp}
In this section, we evaluate the performance of our algorithms by demonstrating the results of our extensive experiments. First, we compare the efficiency of our \lrg{F}$AST$-\lrg{E}$ND$-\lrg{S}$TOPS$ approach with that of the \lrg{B}$ASELINE$-\lrg{E}$ND$-\lrg{S}$TOPS$ algorithm for our $OES$ query. Second, we provide a similar comparative analysis between the \lrg{O}$PT$-\lrg{S}$TOPS$ and the \lrg{H}$EUR$-\lrg{S}$TOPS$ algorithms for the $ORIS$ query. Last, we show the accuracy and scalability of our \lrg{H}$EUR$-\lrg{S}$TOPS$ approach.

We have conducted our experiments in PowerEdge R820 rack server with 6-core Intel Xeon processor, E5-4600 product family, and 64 GB of main memory (RAM). We have compiled our implementations by $GNU\ G++$ with $-O3$ command line option in UNIX OS. In the experiments, we have used a road network graph of San Francisco, CA, USA  \cite{DBLP:journals/geoinformatica/Brinkhoff02}, with $174956$ nodes and $223001$ bi-directional edges. We have loaded the map data and all the necessary data structures into the main memory.

\subsection{The Optimal End-Stops}
\label{a1}
We have artificially generated a set of queries for our $OES$ problem as follows. First, we calculate the maximum node-to-node Euclidean distance, $ED$. Then, we find a pair of nodes, the Euclidean distance between which is approximately a variable percentage of $ED$.  We consider these two nodes as rough estimates of the source-and-destination-cluster-centers. Next, we choose a square window - centered on each estimated cluster center - with its area being a variable percentage of an approximate total area of the graph. Finally, we generate each query by randomly picking two nodes - one from each cluster window. We show a list of the parameters along with their ranges, step sizes, and default values in Table~\ref{tab1}.

\begin{table}[!tb]
\centering
\caption{A List of Parameters for the OES Problem.}
\begin{small}
\begin{tabular}{|c|c|c|c|} \hline
Parameter&Range&Step Size&Default Value\\ \hline
Cluster Distance&30\%-90\%&15\%&60\%\\
(\% of Max Distance)&&&\\ \hline
Cluster Area&1\%-13\%&3\%&7\%\\
(\% of Total Area)&&&\\ \hline
Number of Queries&10-50&10&30\\ \hline
\end{tabular}
\end{small}
\label{tab1}
\end{table}

\begin{figure}[!tb]
\centering
\includegraphics[width=\columnwidth]{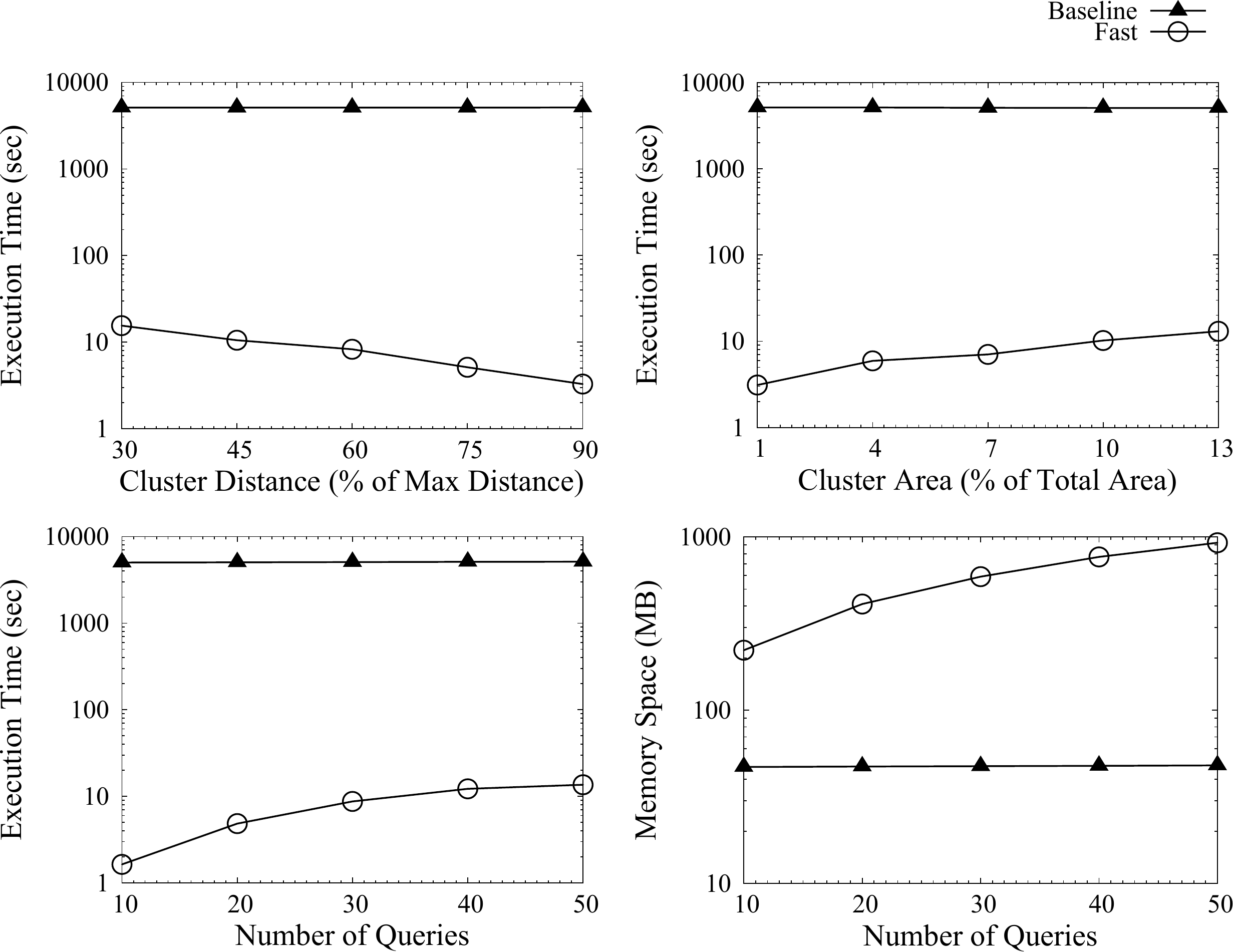}
\caption{Execution Time, and Space - Baseline vs. Fast.}
\label{fig_a1}
\end{figure}

We vary each parameter within its range by its step size, keeping the others at their default values. For each combination of parameter values, we perform $10$ experiments and take an average of the results of these experiments for each of our performance measures, namely, the \emph{execution time}, and the \emph{memory space}. We illustrate our findings in Figure~\ref{fig_a1}. From the figure, it is apparent that our approach performs much better than the baseline brute-force technique in terms of both time and space. Below, we discuss the effect of varying each parameter separately.

\subsubsection{Varying the Cluster Distance}
\label{vcd}
We vary the Euclidean distance between the approximate cluster centers as a variable percentage of the maximum node-to-node Euclidean distance in the graph. This parameter does not affect the \emph{execution time} of the baseline algorithm. Contrarily, our fast solution method shows a negative correlation to the \emph{cluster distance}. This correlation is a consequence of the path-coherence property of road networks. With a fixed \emph{cluster area}, an increase in the \emph{cluster distance} results in a larger amount of shared paths among the passengers. When the users share more route in common, our solution technique relaxes the queries along those routes simultaneously. This triggers a boost in the efficiency. The \emph{memory space} measure does not correlate to the \emph{cluster distance} parameter.

\subsubsection{Varying the Cluster Area}
\label{vca}
We vary the area of each cluster window as a variable percentage of the total area of the road network. Like our analysis in Section~\ref{vcd}, with a fixed \emph{cluster distance}, a larger \emph{cluster area} causes a decrease in the amount of shared routes among the users. Therefore, the \emph{execution time} of our fast algorithm demonstrates a positive correlation with the \emph{cluster area} parameter. On the contrary, the \emph{execution time} of the baseline method remains fixed. This parameter does not affect the \emph{memory space} requirement of either algorithm.

\subsubsection{Varying the Number of Queries}
\label{vnq}
As the complexity analysis in Section~\ref{our_complexity} suggests, both the \emph{execution time} and the \emph{memory space} of our fast solution approach increase linearly in the \emph{number of queries}. However, the measures for the baseline approach remain almost constant, as its complexity analysis in Section~\ref{baseline_complexity} indicates.

\subsection{The Optimal Route and Intermediate Stops}
We have produced a set of synthetic queries for the $ORIS$ problem as follows. First, we find a pair of nodes, whose Euclidean distance is a variable percentage of the maximum node-to-node Euclidean distance in the road network. We regard these two nodes as the end-stoppages of the vehicle. Then, we compute an ellipse as our query space, with the start-and-end-stops of the vehicle as the foci, and an area, a variable percentage of the total area of the graph. Finally, we randomly select the query nodes, for our second problem, from within that space. In Table~\ref{tab2}, we present a list of the parameters along with their ranges, step sizes, and default values.

\begin{table}[!tb]
\centering
\caption{A List of Parameters for the ORIS Problem.}
\begin{small}
\begin{tabular}{|c|c|c|c|} \hline
Parameter&Range&Step Size&Default Value\\ \hline
Euclidean Distance&30\%-90\%&15\%&75\%\\
(\% of Max Distance)&&&\\ \hline
Query Space&10\%-90\%&20\%&50\%\\
(\% of Total Area)&&&\\ \hline
Number of Queries&10-50&10&30\\ \hline
$R_{1}$&.001\%-10\%&10&$\infty$\\
(\% of Shortest Route Cost)&&(Multiply)&\\ \hline
$R_{5}$&0.4-0.8&0.1&0.5\\ \hline
\end{tabular}
\end{small}
\label{tab2}
\end{table}

\begin{figure}[!tb]
\centering
\includegraphics[width=\columnwidth]{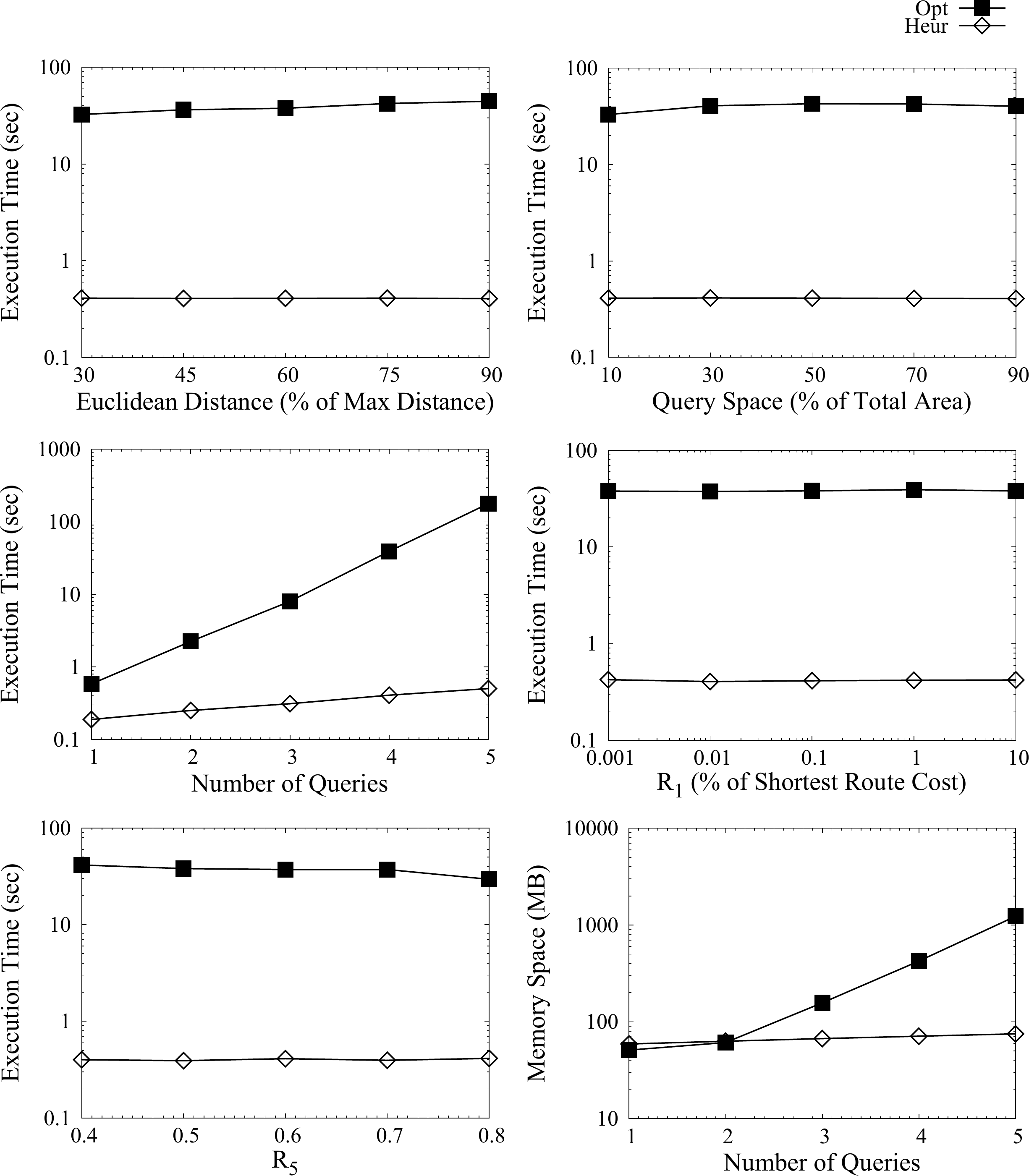}
\caption{Execution Time, and Space - Optimal vs. Heuristic.}
\label{fig_a2}
\end{figure}

\begin{figure}[!tb]
\centering
\includegraphics[width=\columnwidth]{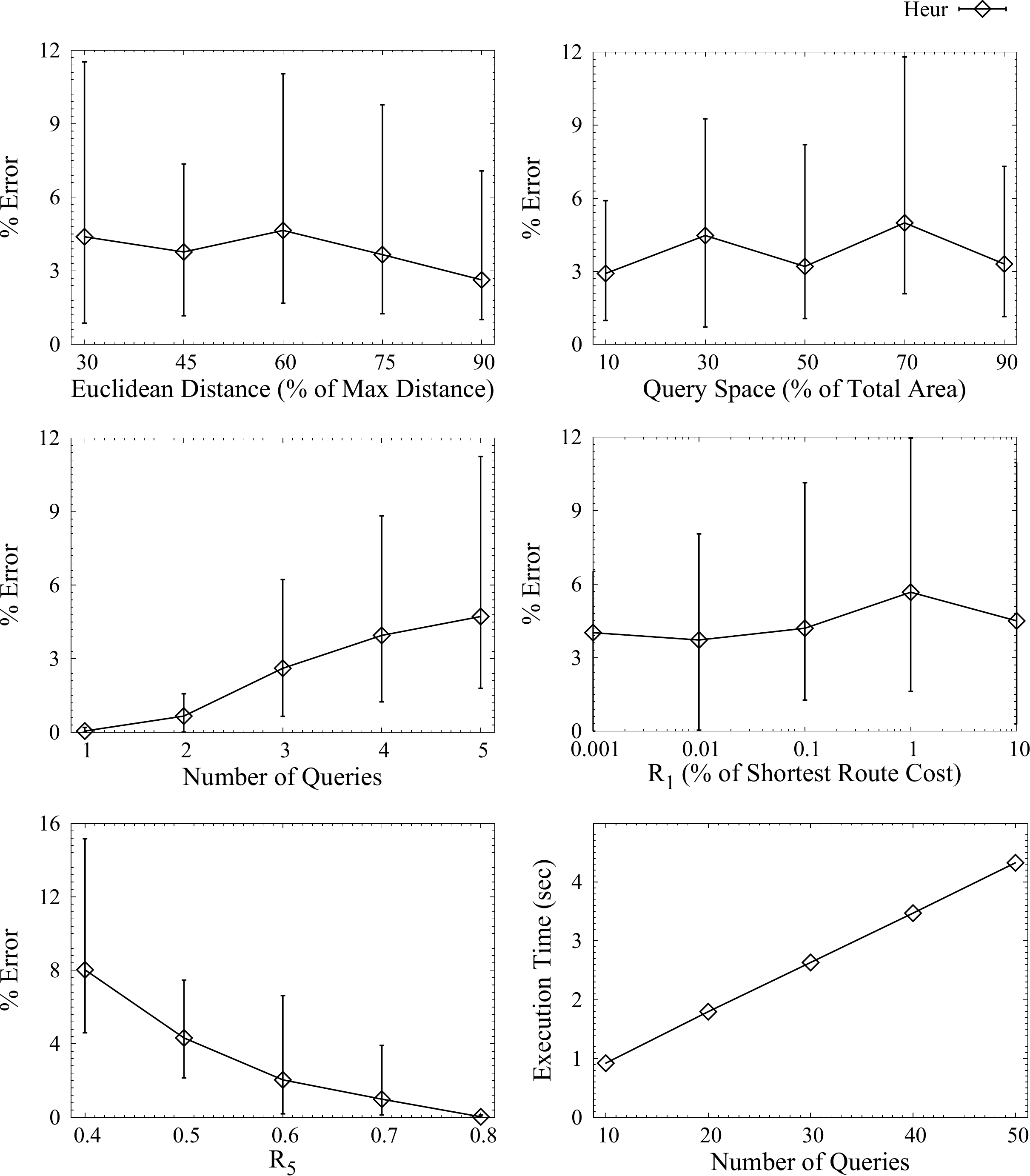}
\caption{Relative Error, and Scalability of Heuristic.}
\label{fig_a22}
\end{figure}

We do not experiment on the second, third, and fourth variants as they would produce results similar to that of the first variant. We produce combinations of parameter values by varying each within its range by its step size, keeping the other parameters at their default values. We carry out $10$ experiments for each collection of parameter values. For the \emph{execution time} and the \emph{memory space} measures, we take an average of the results of these experiments. We show the results in Figure~\ref{fig_a2}. We also measure the minimum, the maximum, and the average \emph{errors} of our heuristic algorithm, which we plot in Figure~\ref{fig_a22}. This figure also demonstrates the \emph{scalability} of the heuristic for a large \emph{number of queries}. Below, we separately investigate the effects of varying each parameter.

\subsubsection{Varying the Euclidean Distance}
\label{ved}
We vary the Euclidean distance between the end-stoppages of the vehicle as a variable percentage of the maximum node-to-node Euclidean distance. The \emph{execution time} of the exact method raises slightly with an increase in this parameter; the pruning method in Section~\ref{imp_a3} can prune more search space, when this parameter is small. However, the \emph{execution time} of the heuristic remains almost the same. This parameter does not affect the \emph{memory space} much. Notice that in Figure~\ref{fig_a22}, the \emph{error} curve fluctuates randomly. This variation is a result of the random orientation of the query nodes and not related to the \emph{Euclidean distance} parameter.

\subsubsection{Varying the Query Space}
\label{vqs}
We vary the area of the query space as a variable percentage of the total area of the graph. As the plots in Figure~\ref{fig_a2} and Figure~\ref{fig_a22} show, none of our performance measures correlates to this parameter. The complexity analyses of our algorithms also support this fact. Like in Section~\ref{ved}, the sway in the \emph{error} curve results from a randomness in the query distribution.

\subsubsection{Varying the Number of Queries}
\label{vnq2}
Figure~\ref{fig_a2} and Figure~\ref{fig_a22} verify our complexity analyses in that the time and the space complexities of the exact algorithm are exponential, while those of our heuristic are linear in the \emph{number of queries}. In figure~\ref{fig_a22}, we have established that the heuristic technique computes an answer within a few seconds, even for a large number of queries. The \emph{error} curve heightens only slightly with an increase in the \emph{number of queries}. However, the average \emph{error} remains quite low. Thus, the gain in efficiency and scalability is worth the little loss of accuracy.

\subsubsection{Varying the Constraint $R_{1}$}
\label{vcr1}
We vary our first constraint $R_{1}$ (Section~\ref{pc1}) as a variable percentage of the shortest path cost between the end-stops. The \emph{execution times}, and the \emph{memory spaces} do not show any correlation to this parameter. The \emph{error} fluctuates randomly as a result of the stochasticity in the location of the query nodes.

\subsubsection{Varying the Constraint $R_{5}$}
\label{vcr5}
The \emph{execution times}, and the \emph{memory spaces} do not vary significantly with a change in the value of our fifth constraint (Section~\ref{pc5}). The \emph{error} falls logarithmically, when we increase the weight of the vehicle's route cost in the objective function. Notice that this behavior is desirable since placing a larger weight in the vehicle's cost prevents looping of its route. For a pragmatic value of $R_{5}$, like $0.75$, the \emph{error} of our heuristic becomes close to zero. Conversely, when $R_{5}$ is close to $0.\overline{3}$, the $TSPP$ case, the \emph{error} of our fifth variant is quite high. However, another way to solve the $TSPP$ is to use our first variant with $R_{1}=0$. Luckily, the \emph{error} of our first variant is quite low for a small value of $R_{1}$. Hence, for a small number of passengers, when approximating the $TSPP$ is more practical, we may use our first variant. When the number of queries is large, we may switch to our fifth variant with a large value of $R_{5}$. This way, we may always compute a pragmatic route with a very low error.

\section{Conclusions}
\label{concl}
We have introduced a new problem of determining the optimal end-stoppages of a vehicle for a group of users. To solve this task, we have proposed a novel algorithm that exploits the path-coherence property of road networks and is many times faster than the baseline brute-force solution. Later, we have presented another new problem of finding the optimal sequence of intermediate stops, given path queries from users and a pair of end-stops. For this, we have outlined an exponential-time-and-space exact algorithm and provided a new polynomial heuristic that efficiently computes a near-optimal solution with a very low error. We have also formulated several variants of this problem and suggested modifications of our algorithms to solve those versions. We have illustrated our techniques with examples and provided detailed complexity analyses.

We have performed extensive experiments, which empirically demonstrate the efficiency and effectiveness of our approaches. In a sample road network with $174956$ nodes and $223001$ bi-directional edges, for a reasonable number of queries, our efficient solution approach for the $OES$ problem is nearly a thousand times faster than the straightforward baseline technique. For the $ORIS$ problem, our experiments visibly demonstrate the polynomial complexity of our heuristic compared to an exponential growth of the exact algorithm. The exact method is intractable for even six or seven queries. For five queries, our heuristic incurs an average relative error of only 5\%. The experiments show that our approach instantly computes a solution for even fifty queries. Thus, our heuristic solution is applicable, when the passengers share a large vehicle such as a mini-bus, or a bus.

As future work, we intend to extend our simultaneous search methodology to solve a variety of new problems. For example, we plan to figure out solutions for the k-optimal pair of end-stops, and the k-optimal sequence of intermediate stops queries, given the same input as in our problems. The motivation behind these new queries is to provide a driver with more choices. Restricting his/her option to a single optimal sequence of stops is often not pragmatic. Certain unpredictable conditions may appear on a road network that may render an optimal route unusable. Again, the optimal path may be loopy, which is impractical. Given some alternatives, a driver may choose the route that best suits him/her.

\balance

\bibliographystyle{ACM-Reference-Format}
\bibliography{sigproc} 

\end{document}